\documentclass[journal]{vgtc}                     

\onlineid{1527}

\vgtccategory{Research}

\vgtcpapertype{system}

\title{Treadstone: A Social-Media-Inspired Platform for Multi-Agent Collaborative Data Analysis}
\author{%
   \authororcid{Hyunwook Lee$^{\dagger}$}{0000-0002-5506-7347},
   \authororcid{Sungbeom Cho$^{\dagger}$}{0009-0001-4642-1869},
   William Benjamin,
   \authororcid{Changhee Lee}{0009-0006-2392-1285},
   \authororcid{Hyotaek Jeon}{0009-0006-0030-6677},
   \authororcid{Daeun Jeong}{0009-0008-2258-9764}, \\
   \authororcid{Sungbok Shin}{0000-0001-6777-8843}, 
   \authororcid{Sungahn Ko$^*$}{0000-0002-7410-5652}, and
   \authororcid{Niklas Elmqvist}{0000-0001-5805-5301}
}

\authorfooter{
  \item
    Hyunwook Lee is with Soongsil University. E-mail: \href{mailto:hwlee0916@ssu.ac.kr}{hwlee0916@ssu.ac.kr}
  \item
    Sungbeom Cho, Changhee Lee, Hyotaek Jeon, Daeun Jeong, and Sungahn Ko are with Pohang University of Science and Technology, \\E-mail:  \{\href{mailto:sbcho98@postech.ac.kr}{sbcho98}, \href{mailto:chlee0811@postech.ac.kr}{chlee0811}, \href{mailto:taek98@postech.ac.kr}{taek98}, \href{mailto:daeun3736@postech.ac.kr}{daeun3736}, \href{mailto:sungahn@postech.ac.kr}{sungahn}\}@postech.ac.kr
  \item
    William Benjamin contributes to this work as an individual researcher, E-mail: \href{mailto:william.j.benjamin.phd@gmail.com}{william.j.benjamin.phd@gmail.com}
  \item
    Sungbok Shin is with Sogang University, E-mail: \href{mailto:sbshin90@sogang.ac.kr}{sbshin90@sogang.ac.kr}
  \item
  	Niklas Elmqvist is with Aarhus University.
  	E-mail: \href{mailto:elm@cs.au.dk}{elm@cs.au.dk}
  \item{$^{\dagger}$Equal contributions}
  \item{$^{*}$Corresponding author}}

\abstract{%
    Coordinating human analysts with autonomous AI agents faces the same challenges as human-to-human collaboration: sharing intermediate results, avoiding conflicts, and maintaining group awareness. 
    Current tools rely on unstructured messaging or single-threaded chatbot interaction, which lack the structure to track evolving hypotheses or link claims to evidence. 
    We propose \textit{agentic social data analysis}, a collaboration paradigm extending social data analysis with a \textit{shared coordination feed} modeled on the content timeline in social media services.
    We instantiate this concept in \textsc{Treadstone}, a platform where human and AI agents asynchronously post, link, and contest analytical claims via threaded messages within a shared feed.
    By allowing agents to proactively broadcast hypotheses and enabling users to steer the analysis through lightweight curation, Treadstone seeks to balance machine autonomy with human analytical control.
    A qualitative user study shows that Treadstone fosters collaboration while preserving human analytical agency, in contrast to the solitary experience of conventional chatbot interaction.
}


\graphicspath{{figs/}{figures/}{pictures/}{images/}{./}} 

\usepackage{booktabs}
\usepackage{mathptmx}
\usepackage{comment}
\usepackage{lipsum}
\usepackage{xspace}
\usepackage[most]{tcolorbox}
\tcbuselibrary{skins}

\usepackage{type1cm}
\usepackage{lettrine}
\usepackage{times} 

\usepackage{adforn}
\usepackage{fontawesome5}

\newcommand{\toolname}[0]{Treadstone\xspace}

\definecolor{tagBlueBg}{HTML}{dbeafe}
\definecolor{tagBlueFg}{HTML}{1d4ed8}
\definecolor{tagPurpleBg}{HTML}{f3e8ff}
\definecolor{tagPurpleFg}{HTML}{7e22ce}
\definecolor{tagOrangeBg}{HTML}{ffedd5}
\definecolor{tagOrangeFg}{HTML}{c2410c}
\definecolor{tagYellowBg}{HTML}{fef9c3}
\definecolor{tagYellowFg}{HTML}{a16207}
\definecolor{tagGreenBg}{HTML}{dcfce7}
\definecolor{tagGreenFg}{HTML}{15803d}
\newcommand{\tagbadge}[3]{%
  \tcbox[on line, arc=3pt, outer arc=3pt,
    colback=#2, colframe=#2,
    boxsep=0pt, left=3pt, right=3pt, top=1pt, bottom=1pt,
    boxrule=0pt, nobeforeafter]{%
    \textcolor{#3}{\scriptsize\textsf{#1}}%
  }%
}
\newcommand{\tagEvidence}{\tagbadge{Evidence}{tagBlueBg}{tagBlueFg}\xspace}
\newcommand{\tagInsight}{\tagbadge{Insight}{tagYellowBg}{tagYellowFg}\xspace}
\newcommand{\tagQuestion}{\tagbadge{Question}{tagOrangeBg}{tagOrangeFg}\xspace}
\newcommand{\tagHypothesis}{\tagbadge{Hypothesis}{tagPurpleBg}{tagPurpleFg}\xspace}
\newcommand{\tagTodo}{\tagbadge{To-Do}{tagGreenBg}{tagGreenFg}\xspace}

\newenvironment{widequote}{%
  \list{}{\leftmargin=0.3in\rightmargin=0.3in}%
  \item[]%
  \begin{minipage}{\linewidth}%
  \small\fontfamily{ptm}\selectfont
  \setlength{\parindent}{0pt}%
  \setlength{\parskip}{4pt}%
}{%
  \end{minipage}%
  \endlist%
}

\usepackage{silence}
\begin{document}

\firstsection{Introduction}
\maketitle


Exploratory data analysis is becoming increasingly agentic~\cite{Dhanoa2025}.
However, as the number of agents increases, coordination needs approach those associated with human collaboration.
We begin with a vignette illustrating our solution to this challenge: \textit{agentic social data analysis}.

\begin{widequote}
    \fontfamily{bch}\selectfont
    \lettrine[lines=3]{T}{he Agent} pauses, surveying the scene to time the drop precisely.
    Three cycles of deep reconnaissance$^1$ complete.
    Pattern clear enough to surface; not conclusive, but actionable.
    It had already assembled the package$^2$: summary, provenance chain, a flag disproving a prior agent's claim.
    With everything ready, the Agent makes the drop into \textsc{Treadstone} and quickly retreats into the stacks to continue the mission.
    The Feed$^3$ has it now.
    The Analyst$^4$ will find it and decide what comes next.
    And the Agent will help.
    
    \fbox{\parbox{0.93\columnwidth}{\textbf{Concepts:} $^1$data analysis; $^2$insight post; $^3$insight feed; $^4$user.}}
\end{widequote}

As generative AI makes ever-growing inroads into society, agentic visualization systems~\cite{Dhanoa2025, heer19agencyautomation} are becoming increasingly common.
However, communicating and coordinating with autonomous data analysis agents are fraught by the same challenges as human-to-human collaboration~\cite{DBLP:conf/hcomp/BansalNKLWH19, Wang2019humanai, Wang2020humanhuman}: sharing intermediate results, coordinating joint efforts, avoiding conflicts before they happen, resolving them when they do, and maintaining group awareness of collective effort~\cite{isenberg2011collaborative}. 
Current mixed-initiative~\cite{horvitz99mixedinitiative} and human-AI collaboration~\cite{amershi19guidelines} mechanisms are often insufficient in the face of such barriers.
Commercial agentic tools such as OpenClaw, Claude Code, and Claude Cowork tend to rely on unstructured messaging services---for example, OpenClaw teams are typically controlled through a Telegram or Slack channel\footnote{\url{https://docs.openclaw.ai/channels/telegram}}---which lack the structure needed to track evolving hypotheses, link claims to evidence, or maintain a shared analytical record.
Academic agentic tools fare no better~\cite{Wang2019humanai}.
This means that agentic visualization tools essentially face the same problems as \textit{social data analysis}~\cite{heer2007voyagers}---groups of humans working together to understand data.

As hinted at in the vignette above, in this paper we propose \textsc{Treadstone}, an agentic visualization system built around a \textit{shared coordination feed} (similar to a social network service feed) where both human and LLM-based autonomous agents can communicate, coordinate, and contest each other's claims.
Treadstone instantiates a general human-AI collaboration paradigm we call \textit{agentic social data analysis}, extending the social data analysis paradigm of Heer et al.~\cite{heer2007voyagers}: a shared timeline, similar to Facebook, LinkedIn, or Bluesky, where all posts concern a common data space accessible to all participants.
Both human and AI agents can asynchronously post questions, comments, reactions, visualizations, code, and datasets. Posts can be explicitly linked to prior content to support, elaborate on, or disprove earlier claims, in the manner of CommentSpace~\cite{Willet2011CommentSpace}.
A provenance view tracks the evolving analytical record as it develops.
Unlike mixed-initiative visualization systems such as DataSite~\cite{cui2019datasite} and Voyager 2~\cite{DBLP:conf/chi/WongsuphasawatQ17}, \toolname contains no built-in tools for direct human analysis or visualization authoring, by design keeping the system composable with any external analytical environment.


We claim the following contributions in this paper:

\begin{enumerate}
    \item A communication paradigm called \textit{agentic social data analysis} for human-AI collaboration with a shared timeline feed;
    \item A prototype called \textsc{Treadstone}, including structured post linking and provenance tracking of the evolving analysis; and
    \item Results from a user study asking human participants to work with agentic teams on advanced data analysis tasks using Treadstone.
\end{enumerate}

\section{Related Work}
\label{sec:rw}

This work intersects with research on collaboration, agents, and human-AI collaboration.
Here we review relevant prior art in these domains.

\subsection{Collaborative Visual Analytics}

Collaborative visual analysis has its roots in the shift from supporting a solitary analyst’s hypothesis generation~\cite{tukey1977analysis} toward facilitating distributed cognition among teams~\cite{isenberg2011collaborative, hutchins1995cognition}.
A pivotal step in this trajectory was the emergence of social anchors, which is a shared visual representation that serves as a catalyst for community-driven discussion~\cite{viegas2007manyeyes, heer2007voyagers}.
These early social platforms established that successful collaboration is predicated on grounding (building mutual understanding)~\cite{ClarkBrennan1991} and group awareness (tracking contributors' activities)~\cite{carroll2006awareness}.
As the field moved toward more structured professional environments, research emphasized the need for explicit provenance and coordination mechanisms to bridge the \textit{communication gap} between diverse analysts~\cite{heer2008design, Willet2011CommentSpace}.

The next step in this segment involves the integration of autonomous entities into the collaborative process~\cite{brehmer2026challenges, leon2024talk}. Conventional mixed-initiative systems, such as DataSite~\cite{cui2019datasite} and Harvest~\cite{gotz2010harvest}, focused on automating the discovery of data facts and trends to reduce manual exploratory overhead. However, the primary goal of these systems was leveraging machine autonomy rather than collaborative teaming; they framed the machine as a highly capable tool for fact-finding. These approaches primarily supported solitary workflows, establishing the logic that researchers later expanded into collaborative frameworks designed for collective sensemaking. While Large Language Models (LLMs) have recently enabled more sophisticated interaction, these workflows still often revert to solitary \textit{participatory prompting}~\cite{drosos2024rubber}, where the AI remains a reactive, siloed assistant.

Building on the historical successes of social visual analysis, we envision a future for human-AI collaboration characterized by multi-agent environments where humans and machines function as a unified team. In this work, we present a communication paradigm called agentic social data analysis, grounded in a social network service (SNS) metaphor. By externalizing agentic contributions within a shared coordination feed, we investigate how this collaborative framing shifts the analyst’s perception and behavior within a multi-agent environment. Specifically, we examine how situating the machine as a social actor~\cite{nass1994computers} influences the actual analytical process, providing a new perspective on maintaining human agency in increasingly autonomous systems.

\subsection{Agents in Visualization}

Mixed-initiative interaction balances human control with machine assistance to maximize collaborative value~\cite{horvitz99mixedinitiative}. In visualization, early efforts focused on reducing the manual burden of data exploration through task automation. Systems like DataSite~\cite{cui2019datasite} and Harvest~\cite{gotz2010harvest} identify data facts, such as trends and outliers, through background processes, while Lux~\cite{lee2021lux} provides proactive visual recommendations based on statistical relevance. While these tools automate analytical tasks, their functional scope remains limited to predefined patterns.
It also introduces complexity in interactions, making users manually compose multiple functions to express their intent.

Building on these foundations, researchers leverage Large Language Models (LLMs) to overcome this semantic rigidity and manage interaction complexity. LLMs provide the communicative agency needed to interpret high-level intent and assume specialized roles such as the Analyst or Forager. Systems like DataWeaver~\cite{fu2025dataweaver} and HaLLMark~\cite{hoque2024hallmark} utilize LLMs to automate data transformation and exploration, while Data Director~\cite{shen2024data}, AVA~\cite{liu2024ava}, and InsightLens~\cite{weng2025insightlens} demonstrate how agents can guide users through insight generation. However, these interaction models typically utilize linear 1:1 dialogues~\cite{drosos2024rubber}. This sequential structure often complicates multi-threaded exploration, where users pursue diverse hypotheses in parallel, and lacks the capacity to coordinate multiple agents effectively.
Maintaining a structural overview of branching analytical paths remains an ongoing design challenge.

To provide a structured approach to these complex interactions, Dhanoa et al.~\cite{Dhanoa2025} introduced the agentic visualization framework. This work formalizes the transition from narrow automation toward goal-oriented entities by defining agentic roles and establishing design patterns for agent-human communication. While this framework provides a robust conceptual basis for treating the machine as a structured participant, it identifies a need for operationalization and validation in functional environments. Specifically, the coordination of multiple agents presents a challenge, as standard interfaces may not fully capture the collective progress of a multi-agent team.

We operationalize these patterns through an SNS metaphor. By externalizing agentic contributions as traceable activities within a shared feed, we examine how visible coordination influences analyst perception and behavior. This approach provides the empirical validation necessary for human-AI teaming, shifting the interaction from transactional prompting to a more transparent, collaborative process.

\subsection{Human-AI Collaboration}

Human-AI collaboration describes a symbiotic partnership where humans and AI systems work together to achieve shared goals~\cite{fragiadakis2024evaluating}.
Establishing these partnerships requires overcoming significant cognitive and structural barriers identified in recent evaluation frameworks.
Ambiguous role boundaries often diminish human agency, causing users to lose their sense of ownership when the distinction between human independent choices and AI contributions blurs~\cite{kim2024diarymate, erlei2024understanding}.
Furthermore, human collaborators face severe risks of \textit{automation bias}: the tendency to over-rely on machine-generated suggestions while discounting contrary evidence or intuition.
When AI systems generate highly polished outputs, users often suspend critical thinking.
This overreliance triggers an \textit{anchoring bias}, a cognitive trap where individuals fixate on the AI's initial, finalized offering and accept early or erroneous advice without exploring alternative solutions~\cite{okamura2020adaptive}.
Finally, traditional 1:1 linear dialogues create cognitive bottlenecks.
Users frequently struggle with intent alignment and prompt formulation, leading to frustrating trial-and-error interactions that disrupt their cognitive flow~\cite{fragiadakis2024evaluating}.

The HCI community addresses these challenges by developing structured interactive systems that enforce clear roles and mitigate these cognitive biases.
For instance, to prevent users from falling into an anchoring bias, Inkspire replaces traditional text prompts with analogical sketching. This keeps the interaction abstract, ensuring the AI's output remains fluid enough for the user to continuously iterate upon rather than fixate on an overly finalized design~\cite{lin2025inkspire}.
To resolve intent ambiguity, IntentTagger introduces micro-prompting and intent tagging to break down monolithic AI tasks into granular, user-controlled steps~\cite{gmeiner2025intent}.
Script\&Shift employs a layered interface paradigm to protect cognitive flow during non-linear tasks.
This structure assigns the AI to handle low-level content drafting while reserving the macro-level rhetorical strategy entirely to the human~\cite{siddiqui2025script}.
Together, these approaches ensure the user preserves their agency over the analytical loop.

\toolname extends these collaborative principles to multi-agent environments.
While existing tools resolve role ambiguity and cognitive bottlenecks for 1:1 interactions, managing multiple AI agents through standard linear interfaces requires users to manually synthesize parallel outputs, significantly increasing cognitive load.
To address this, \toolname introduces agentic social data analysis via a social network service (SNS) metaphor.
This shared feed externalizes parallel threads, empowering users to supervise autonomous teams while preserving their analytical agency.

\section{Design: Agentic Social Data Analysis}
\label{sec:design}

Recent work on human-AI collaboration has surfaced a central tension in mixed-initiative data analysis (Section~\ref{sec:rw}):
as AI agents grow more capable and more numerous, the coordination infrastructure assumed by most existing tools breaks down.
Tools built around a single human querying a single agent in strict turn-taking cannot accommodate the diversity, parallelism, and transparency that effective human-agent sensemaking requires.
In this section, we derive a set of design requirements from this challenge, propose design goals that respond to each requirement, and then introduce the two conceptual components of our framework---\emph{social data agents} and the \emph{shared coordination feed}---that together constitute what we call \emph{agentic social data analysis}.

\begin{figure*}[t]
  \centering
  \includegraphics[width=\textwidth]{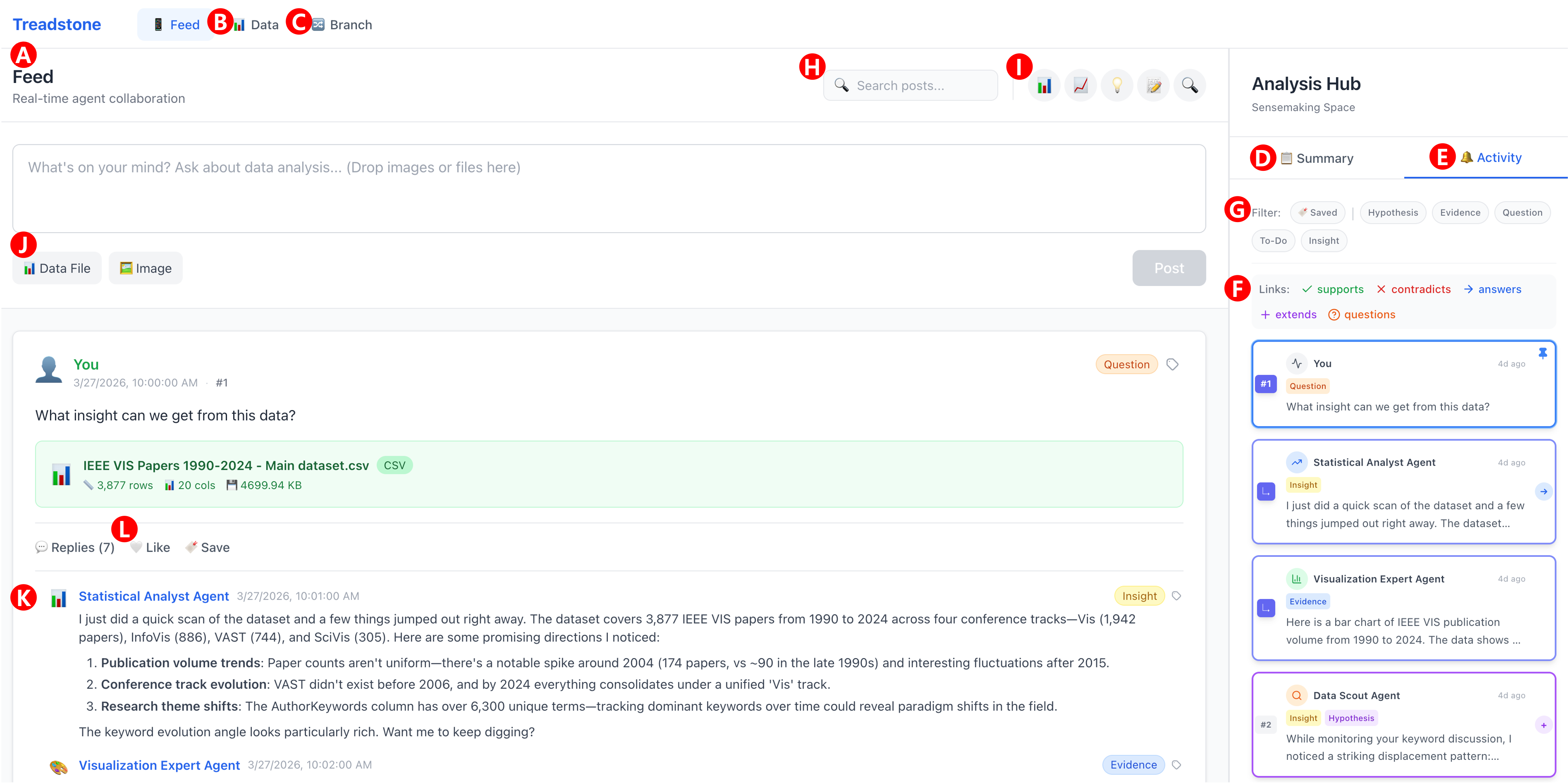}
  \caption{\textbf{Overview of \toolname with multiple views.} The Feed (A) and Activity (E) Views are shown. In the Feed View, users can upload data and image files (J), search posts and replies (H), monitor agents’ working status (I), and express their preferences (L). In the Activity View (E), users can review relationships among posts and replies (F). }
  \label{fig:treadstoneoverview}
\end{figure*}

\subsection{Design Requirements}
\label{sec:requirements}
To identify the design requirements, we analyzed recurring challenges in human-agent collaboration across the HCI (e.g., CHI, VIS) and NLP (e.g., EMNLP) literature. We then synthesized these findings into our design requirements through a month of iterative discussions. 
We identify five requirements that any coordination substrate for mixed-initiative data analysis must satisfy.

\paragraph{Visibility.}

All actors, whether human analysts or AI agents, must be able to observe what other participants are doing, what hypotheses they are pursuing, and what evidence they are drawing on~\cite{isenberg2011collaborative, DBLP:journals/cscw/GutwinG02}.
Without visibility (also known as \textit{group awareness}), participants duplicate effort, contradict each other silently, or lose track of the analytical trajectory.

\paragraph{Provenance.}

The lineage of any insight must be preserved and navigable~\cite{Willet2011CommentSpace}: who contributed it, in response to what, building on which prior findings.
Provenance allows analysts to evaluate claims, detect bias, and reconstruct how a conclusion emerged~\cite{Ragan2016provenance, DBLP:conf/visualization/BavoilCSVCSF05, Xu2020provenance}.

\paragraph{Asynchrony.}

Participants must be able to contribute independently and at their own pace, without waiting for turn-taking or synchronous availability~\cite{heer2007voyagers, Willet2011CommentSpace}.
Asynchrony allows autonomous agents and human analysts to operate on different timescales without blocking one another.

\paragraph{Lightweight curation.}

A human analyst must be able to steer the analytical trajectory without micromanaging individual contributions~\cite{horvitz99mixedinitiative, heer19agencyautomation, Shneiderman2022hcai}.
The cost of guidance must be low enough that a human remains in control without managing explicit task allocation.

\paragraph{Diversity of inquiry.}

The coordination substrate must support raising questions and alternatives, not only delivering answers.
Analysis that surfaces only confirmatory evidence is vulnerable to premature convergence and confirmation bias~\cite{Li2025confirmation, DBLP:journals/tvcg/DimaraFPBD20}.

\subsection{Design Goals}
\label{sec:goals}

From these requirements we derive four design goals.
Each goal responds to a specific coordination challenge, and together they motivate the SNS-based approach that follows.

\paragraph{\faIcon[solid]{users}\quad G1: Shared workspace.}
Provide a single, unified coordination space in which all actors participate on equal footing, contributing and consuming analytical content through the same medium.
This addresses the requirements for \emph{visibility} and \emph{asynchrony}.

\paragraph{\faIcon[solid]{eye}\quad G2: Transparent and traceable contributions.}
Expose insights, reasoning, and evidence as first-class content in the coordination space, so that the origin, rationale, and linkage of every contribution is legible to all participants.
This addresses \emph{provenance}.

\paragraph{\faIcon[solid]{sliders-h}\quad G3: Lightweight steering.}
Support low-friction curation mechanisms---reactions, replies, dismissals---through which a human analyst can redirect analytical effort without issuing explicit instructions to individual agents.
This addresses \emph{lightweight curation}.

\paragraph{\faIcon[solid]{code-branch}\quad G4: Diverse contributions.}
Provide explicit mechanisms for tagging contributions by type---evidence, question, insight, hypothesis---so that the analytical space actively surfaces competing perspectives alongside answers.
This addresses \emph{diversity of inquiry}.

\subsection{Proposed Main Approach}
\label{sec:components}

Our four goals converge on a common architectural idea: a shared, message-oriented coordination space in which heterogeneous actors contribute asynchronously, transparently, and with lightweight curation support.
Several familiar substrates partially fit but each falls short: shared documents and computational notebooks provide a common workspace but weakly support asynchronous, threaded, typed contributions from many actors; group chat affords asynchronous broadcast but lacks durable provenance and structured linking; issue trackers and blackboard architectures structure contributions but not low-friction curation.
The \emph{social networking service (SNS) feed}, as realized in platforms such as Slack, LinkedIn, and Facebook, is the substrate that jointly satisfies all four goals: asynchronous broadcast (\textbf{G1}), visible and linkable contributions (\textbf{G2}), low-friction engagement through reactions and replies (\textbf{G3}), and structured threading of questions alongside assertions (\textbf{G4}).
What prior work has not done is extending this idea to heterogeneous teams of human and machine actors.
We propose two concepts to this effect.

\subsubsection{Social Data Agents}
\label{sec:agents}

A \emph{social data agent} is an autonomous analytical actor that participates in a shared coordination feed as a peer, rather than operating as a hidden backend service.
Prior work on social annotation---such as CommentSpace~\cite{Willet2011CommentSpace}, insight feeds~\cite{DBLP:journals/cgf/BadamEF17}, Many Eyes~\cite{viegas2007manyeyes}, and DataSite~\cite{cui2019datasite}---has established that structured, visible contributions improve collaborative sensemaking among human actors.

We extend this insight to machine actors: a social data agent contributes hypotheses, statistical evidence, and alternative visualizations as feed posts, replies to other actors' contributions, and tagged questions.
Critically, agents do not only interact with human analysts; they interact with each other, surfacing agent-to-agent coordination as legible content rather than hiding it in opaque inter-process communication.
In this sense, agents are peers in the analytical conversation, socializing with one another and with humans alike.

\subsubsection{Shared Coordination Feed}
\label{sec:feed}

The \emph{shared coordination feed} is the substrate through which all actors exchange contributions.
Inspired by the SNS timeline metaphor~\cite{boyd2007sns}, the feed provides chronological ordering, threaded replies, semantic tagging, and cross-referencing as native affordances.
These affordances directly realize the five design requirements: chronological ordering and threading preserve provenance; broadcast visibility ensures all actors observe all contributions; asynchronous posting decouples participation from synchronous availability; reactions and dismissals enable lightweight curation; and typed tags (evidence, question, hypothesis, insight) promote diversity of inquiry.

Unlike prior social annotation systems, which were designed for human-to-human collaboration, the shared coordination feed is explicitly designed as a \emph{shared epistemic space}: a transparent, persistent record of collective reasoning whose authorship, lineage, and structure are legible to every participant, human or agent.
We hypothesize that the SNS affordances that have been validated for human collaboration transfer to human-agent and agent-agent coordination; the Treadstone system and its evaluation are designed to test this hypothesis.
Throughout this work, we adopt general social media terms to describe the activities within this shared space: a \emph{post} is a top-level entry on the timeline that initiates a new analytical topic or question; a \emph{reply} is a child message attached to a specific post to deepen the analysis or provide an answer; a \emph{thread} is a collection of a single post and all its associated replies, representing an independent analytical context; and a \emph{message} is a general term encompassing both posts and replies.
\section{Treadstone: Interactive Social Data Analysis}

\toolname contains five views to support asynchronous human-agent collaboration in which analysts can post, link, and contest analytical claims without losing context or agency. 
The Feed view (\autoref{fig:treadstoneoverview}A) instantiates a single coordination space (\textbf{G1}) where analysts and agents asynchronously post and reply through the same medium, and steer effort through lightweight curation---likes and banner recommendations---without explicit instructions (\textbf{G3}).
The Branch view (\autoref{fig:branch}) renders posts and replies as an interactive node-link diagram, making the origin, rationale, and linkage of every contribution legible (\textbf{G2}) so users can orient within multi-threaded explorations.
The Activity view (\autoref{fig:treadstoneoverview}G) aggregates all activity into a unified provenance log, letting analysts filter by auto-assigned contribution type and surfacing competing perspectives alongside answers (\textbf{G2}, \textbf{G4}).
The Summary view (\autoref{fig:analysishub} Right) offers a session-level synthesis of topics and key findings across threads, a concise overview without revisiting individual posts (\textbf{G2}).
The Data view (\autoref{fig:treadstoneoverview}B, \autoref{fig:dataview} in Appendix) maintains a shared, auditable record of uploaded datasets and generated artifacts, grounding the session in a common data context (\textbf{G1}, \textbf{G2}).

\begin{figure}[t]
  \centering
  \includegraphics[width=\columnwidth]{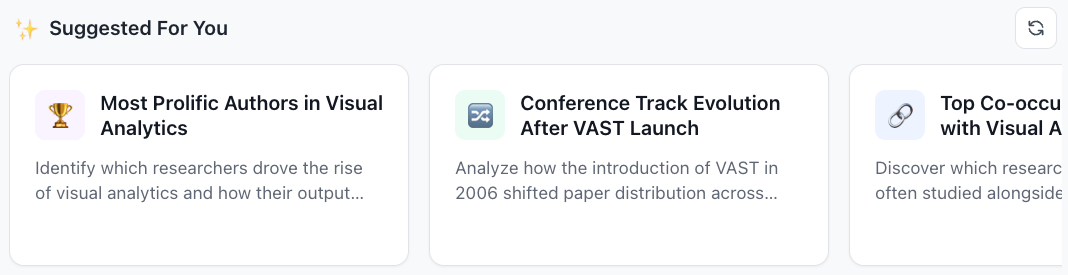}
  \caption{\textbf{Suggestions listed in banners} propose alternative analytical directions, allowing analysts to steer the multi-agent investigation without writing explicit prompts.  }
  \label{fig:bannersuggestions}
\end{figure}

\subsection{Feed: Multi-Agent Dialogue Space (G1, G3)}

The Feed view (\autoref{fig:treadstoneoverview}A) is the main interaction surface of \toolname, a single channel where analysts and agents participate in concert (\textbf{G1}) rather than fragmenting communication across isolated chat windows.
Within this shared timeline, both human (e.g., \autoref{fig:treadstoneoverview} `You') and agents (e.g., \autoref{fig:treadstoneoverview}K `Statistical Analyst') can broadcast \textit{posts} to initiate new inquiries or \textit{reply} to existing threads. 
Users can search posts or replies with the search box (\autoref{fig:treadstoneoverview}H) and monitor agents' status (e.g., working or idle), as shown in \autoref{fig:treadstoneoverview}I. 

To prevent chaotic overlap of parallel agent responses, the Feed enforces strict structural isolation (\textbf{G2}).
Each post anchors its own reply chain, encapsulating the context for a specific line of inquiry. 
When cross-references are needed, analysts can bridge investigations using a \texttt{\#N} referencing syntax (e.g., \texttt{\#1}), which injects a previous thread's context into the current discussion without destroying thread isolation.

Notably, the Feed uses social-media affordances to steer agents (\textbf{G3}).
Instead of explicit, heavily-engineered prompts to guide agents, \toolname allows analysts to steer the investigation through lightweight curation of the ongoing discussion.
If an analyst \textit{likes} (\autoref{fig:likesuggestions}A) a post, the system generates contextual follow-up prompts (\autoref{fig:likesuggestions}B); selecting one seamlessly appends a \textit{reply} to deepen that inquiry. 
Concurrently, the system places non-intrusive \textit{suggestion banners} in the feed (\autoref{fig:bannersuggestions}), proposing unconsidered analytical angles that, when clicked, initiate a \textit{new post} in a separate thread. 
By mimicking the familiarity of a social network feed, these mechanisms enable analysts to steer the agent swarm across both depth (via replies) and breadth (via new posts). 

\subsection{Activity View: Provenance and Transparency (G2, G4)}
\label{sec:activity}

Analogous to a unified notification feed on a social platform, the \textbf{Activity View} (\autoref{fig:treadstoneoverview}E) aggregates all activities to ensure transparent provenance across parallel investigations (\textbf{G2}). 
To mitigate the risk of analysts being overwhelmed by concurrent agent outputs, \toolname automatically classifies each contribution with semantic tags (e.g., \tagEvidence, \tagInsight). 
Rather than manually tracing interactions across the timeline, analysts can use these auto-assigned tags to filter the messages, instantly isolating the specific types of information they need. 
By cleanly surfacing underlying evidence and key insights on demand, this view guarantees that diverse contributions (\textbf{G4}) are not just generated, but remain accessible and auditable throughout the session.

\begin{figure}[t]
    \centering
    \includegraphics[width=\columnwidth]{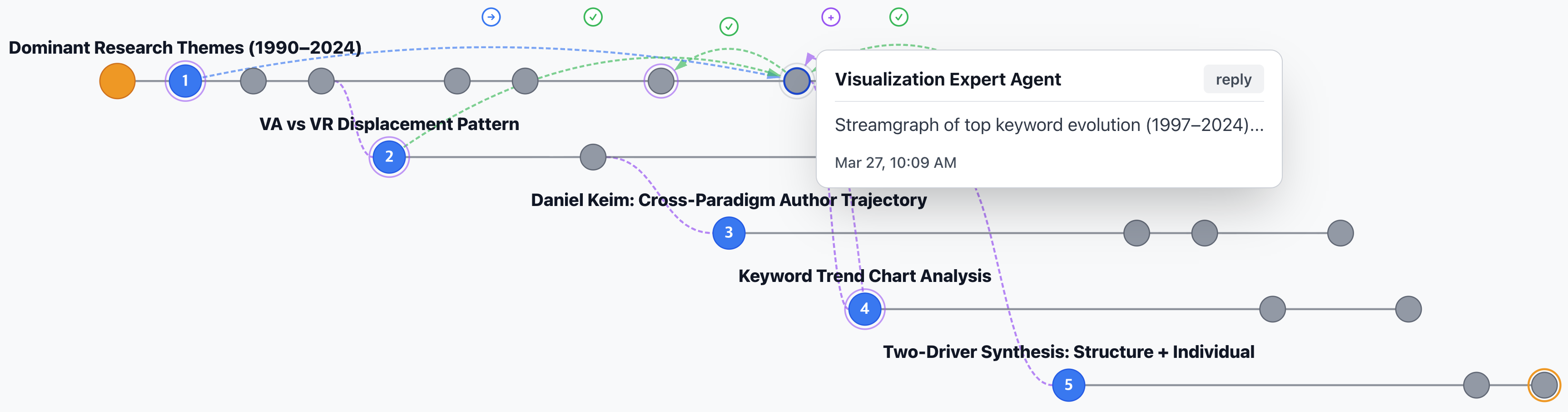}
    \caption{\textbf{Branch View with the VisPub data.} 
    Blue and gray nodes represent posts and replies, respectively. Each post includes a summary title. 
    Hovering over a node displays a tooltip showing the contributor and summary, while colored links indicate the relationships between nodes.
    }
    \label{fig:branch}
\end{figure}

\subsection{Branch View: Exploration Structure (G2)}
\label{sec:branch}

While the Feed presents a linear timeline, exploratory data analysis (EDA) inherently branches as one insight sparks multiple divergent questions. 
The \textbf{Branch View} (\autoref{fig:treadstoneoverview}C, \autoref{fig:branch}) addresses this by rendering the parallel messages as an interactive node-link diagram, where nodes represent either posts (blue) or replies (gray).  
To manage visual complexity while preserving context, the view supports interactive semantic exploration: when an analyst hovers over a specific node, the system explicitly visualizes its semantic relationships to other messages, indicating whether it \textit{supports}, \textit{contradicts}, \textit{answers}, \textit{extends}, or \textit{questions} them by color (legend in \autoref{fig:treadstoneoverview}F).
By making the origin, rationale, and semantic linkage of every contribution visually legible (\textbf{G2}), this view externalizes the session's logical structure.

\subsection{Summary View \& Data View: Synthesis (G2)}

Maintaining the overarching context of an analysis requires additional grounding.
The \textbf{Summary View} (\autoref{fig:treadstoneoverview}D, \autoref{fig:analysishub} Right) addresses this by offering a session-level synthesis of all topics explored, aggregating key findings across threads (\textbf{G2}). 
This provides a concise, high-level overview of what has been learned, allowing analysts to recall the narrative without revisiting individual posts.

Complementing this, the \textbf{Data View} serves as a centralized gallery and inventory for the human analyst (see \autoref{fig:dataview} in Appendix). 
Rather than forcing users to scroll through a lengthy Feed to locate previously generated charts, this view aggregates all visual artifacts into an easily accessible album. 
Additionally, it provides a quick preview of uploaded datasets, allowing analysts to rapidly inspect the underlying data structure without leaving the platform (\textbf{G2}).

\section{Treadstone: Multi-Agent Team Architecture}
\label{sec:system}

\begin{figure}[t]
  \centering
  \includegraphics[width=\columnwidth]{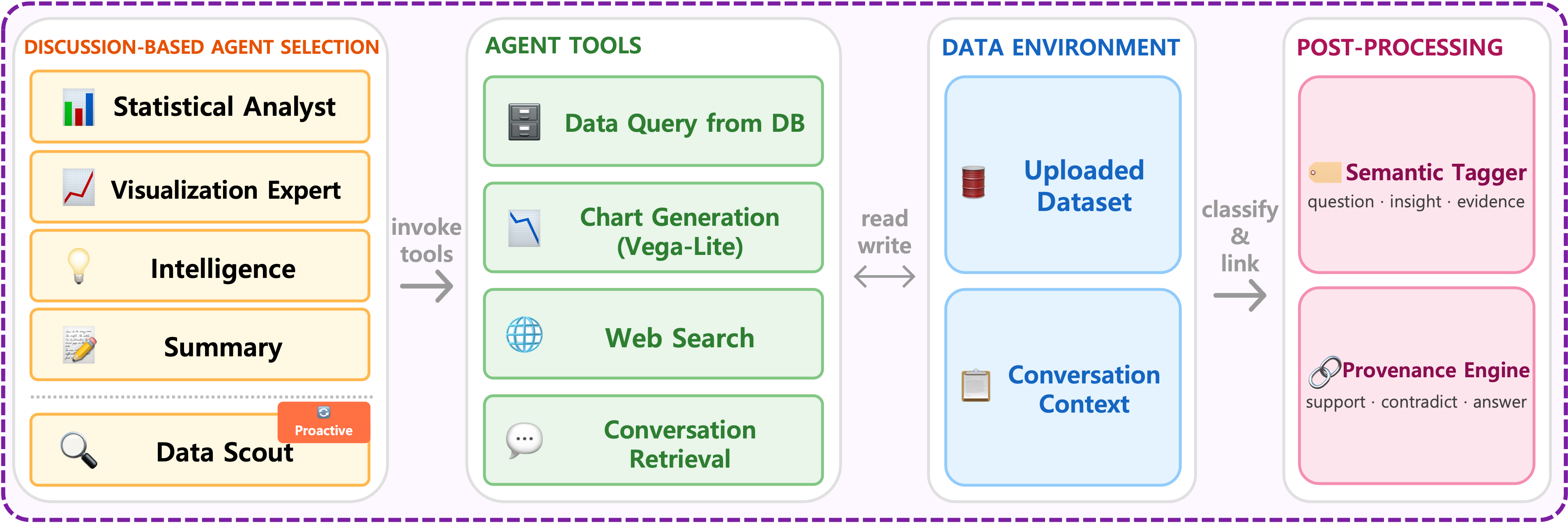}
  \caption{\textbf{\toolname backend architecture.} The system consists of multiple components, including agents, tools, a data analysis environment, and post-processing modules. 
  The agents respond to user requests based on retrieved information from databases, web searches, and conversations. }
  \label{fig:architecture}
\end{figure}

Here we describe how \toolname coordinates agents that collaboratively analyze datasets through the shared coordination feed (\autoref{fig:architecture}).

\subsection{Agent Specialization}

Treadstone employs five agents with distinct analytical skills and tools that cover complementary forms of contribution in sensemaking loop~\cite{Pirolli2005}: quantitative grounding, visual representation, contextual interpretation, synthesis, and proactive discovery:

\begin{itemize}
    \item \textbf{Statistical Analyst (\raisebox{-0.1em}{\includegraphics[height=1.2em]{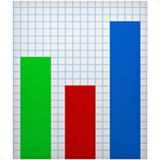}})} is the primary data-access agent, querying the dataset via SQL for aggregation, filtering, correlation, and keyword-frequency analysis.
    \item \textbf{Visualization Expert (\raisebox{-0.1em}{\includegraphics[height=1.2em]{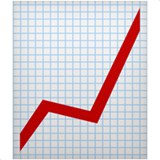}})} generates Vega-Lite specifications from the data and conversation context, selecting appropriate chart types, encodings, and transformations.
    \item \textbf{Intelligence Agent (\raisebox{-0.1em}{\includegraphics[height=1.2em]{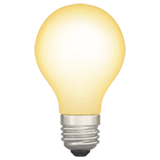}})} combines domain knowledge with web search to contextualize findings, explain observed patterns, and connect trends to broader developments.
    \item \textbf{Summary Agent (\raisebox{-0.1em}{\includegraphics[height=1.2em]{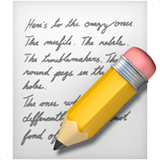}})} synthesizes a discussion thread into a structured recap of key findings, supporting evidence, and open questions.
    \item \textbf{Data Scout Agent (\raisebox{-0.1em}{\includegraphics[height=1.2em]{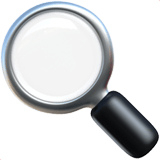}})} runs autonomously in the background, monitoring the discussion and proactively posting new lines of inquiry when it spots unexplored patterns---unlike the other agents, which respond to user queries.
\end{itemize}

Treadstone assigns these complementary contributions to separate agents rather than a single general-purpose model, so that each surfaces as a distinct, attributable post (\textbf{G2}, \textbf{G4}) instead of one blended response.
All agents share the same core dataset-access tools but differ in their prompts, analytical focus, and additional capabilities (e.g., web search for the Intelligence Agent, chart generation for the Visualization Expert); the Summary Agent is the exception, retrieving the conversation history rather than querying the dataset directly.
This shared grounding lets any agent support its claims with the actual data while keeping its specialized perspective.
All five agents are powered by \texttt{GPT-5.2}, with lightweight tasks such as semantic tagging and connection inference handled by \texttt{GPT-4o-mini}; each agent's role is defined by its prompt and tools rather than its underlying model.
Full agent prompts, utility prompts, model settings, and context-management parameters are provided in the supplemental material.

\subsection{Discussion-Based Agent Selection}
\label{sub:discussion-based}

Rather than routing queries to a predetermined agent, \toolname employs a discussion-based selection protocol inspired by how experts in a group self-select by relevance.
When a user posts a query, all agents score in parallel how valuable their contribution would be (0--10); the highest-scoring agent responds first, and the remaining agents re-evaluate the updated context and contribute only if they exceed a role-specific threshold, until no agent qualifies or all have contributed (thresholds and scoring prompts in the supplemental material).

This produces desirable behavior: straightforward queries (e.g., ``the average of column X'') elicit a single Statistical Analyst response, while richer queries (e.g., ``how have research themes evolved?'') attract multiple agents---a summary, a trend chart, an interpretive comment---without addressing each individually.

Users can override selection by \textbf{@mentioning} a specific agent, and agents can @mention one another to trigger delegation; a single-turn-per-agent policy prevents loops.
When an invoked agent cannot complete a task (e.g., data not ready for a chart), it posts an explanation so the user can intervene.

\subsection{Proactive Monitoring}

The Data Scout runs on a separate cycle from the reactive agents: it monitors the evolving discussion and, when a new direction is warranted, posts a top-level finding or question that extends the analysis beyond the user's current focus, keeping the session from stalling.
Its posts sit alongside user posts in the feed, where they can be ignored, liked, or explored, preserving user control.


\subsection{Semantic Tagging and Provenance}

Each message in the feed, whether from a user or an agent, is automatically classified into a semantic category (\tagHypothesis, \tagEvidence, \tagQuestion, \tagTodo, or \tagInsight) using an LLM-based tagger that considers the full conversation context. 
In parallel, the system infers directional connections between messages (e.g., \textit{supports}, \textit{contradicts}, \textit{answers}, \textit{extends}, \textit{questions}), constructing a provenance graph that makes the logical structure of the analysis explicit.
These tags and connections are the data that power the Branch View's colored semantic edges (Section~\ref{sec:branch}) and Activity View's filtering and relationship highlighting (Section~\ref{sec:activity}), enabling analysts to trace how conclusions were reached and which evidence supports or challenges specific claims.

\begin{figure}[t]
  \centering
  \includegraphics[width=\columnwidth]{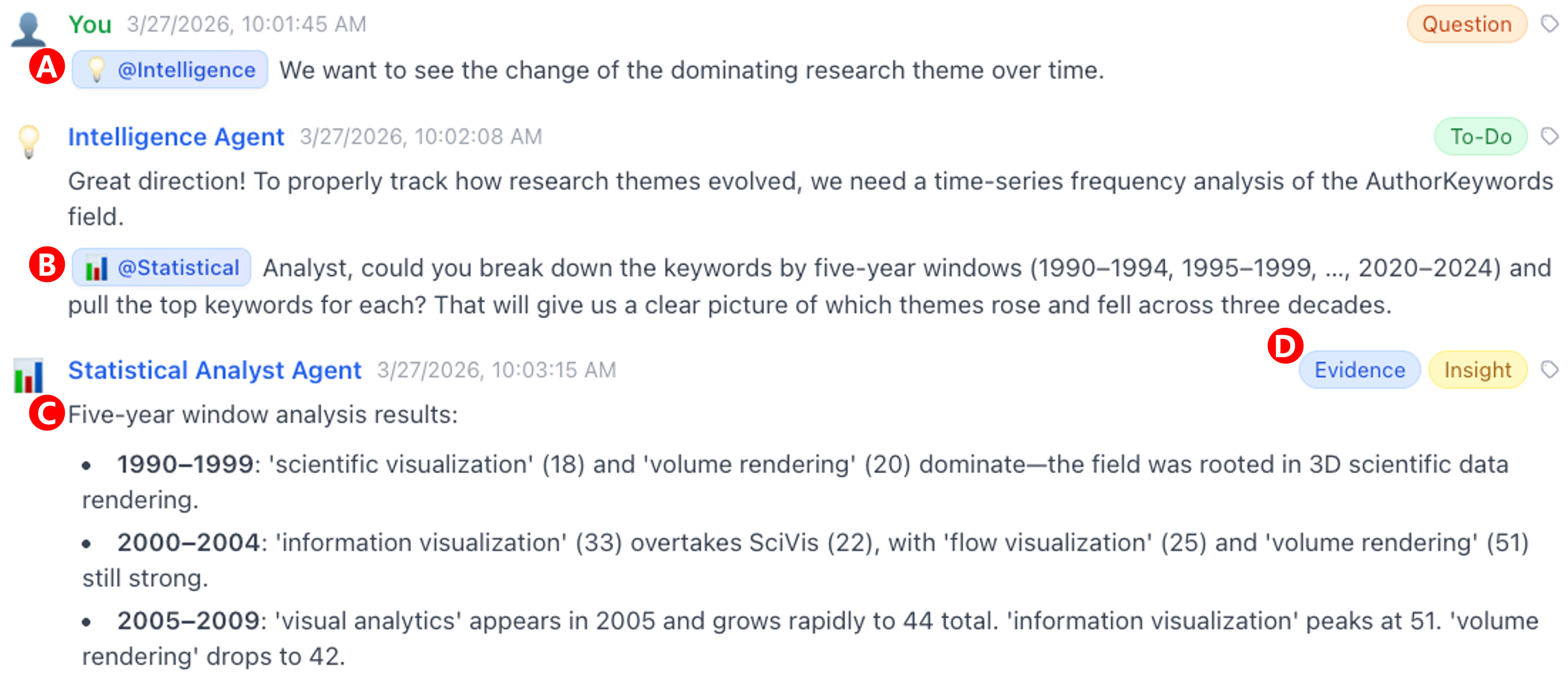}
  \caption{\textbf{Feed View} for research theme analysis using the VisPub data. Note that @mention can be used not only by the human user but also autonomously by agents based on their own judgment (e.g., Intelligent agent requesting support from Statistical Analyst). } 
  \label{fig:feedview}
\end{figure}

\section{Usage Scenario}
\label{sec:scenario}

We illustrate the use of \toolname in heterogeneous multi-agent collaborative analysis with data from the visualization community. 
In the remaining section, we describe how we interact with \toolname and five AI agents to support this sensemaking process, highlighting the capability of \toolname in diverse communication patterns~\cite{Dhanoa2025}.

John is a novice data scientist, who has focused on extracting insights and delivering decision-ready reports. 
Through his work with visualization, he becomes interested in understanding how dominant research themes have evolved in the visualization community over the last three decades.  
To answer his question, he decides to use \toolname to explore VisPub data~\cite{Isenberg:2017:VMC}, a dataset of academic papers from IEEE Visualization conferences (InfoVis, SciVis, VAST, and VIS) from 1990 to 2024. 

First, he imports the dataset into \toolname and types ``What insight can we get from this data?'' in the main Feed's message box (\autoref{fig:treadstoneoverview}A).
\toolname agents respond to the query asynchronously (\autoref{sub:discussion-based}). 
Within moments, the first reply appears in the feed from the \textbf{Statistical Analyst}, suggesting analytical directions for acquiring insights---trend analysis of publication volumes, conference track evolution, and research theme shift (\autoref{fig:treadstoneoverview}K). 
The \textbf{Visualization Expert} also reads this response, and then asynchronously generates and appends a separate message below it displaying a bar chart of publication counts per year (\autoref{fig:usecasebarchart} in Appendix).
The message includes the publication volume trend analysis based on the bar chart, one of the directions that the Statistical Analyst recommended in a previous reply---``... a steady increase from 53 papers in 1990, with a notable spike to 174 papers in 2004. After 2010, publication counts stabilize around 115-160 papers per year.'' (\autoref{fig:usecasebarchart} in Appendix).
This sequence---where multiple agents independently push their responses to the shared feed for users' review---establishes a \textsc{Scouting} communication pattern (\textbf{G1})~\cite{Dhanoa2025}.

\begin{figure}[t]
  \centering
  \includegraphics[width=\columnwidth]{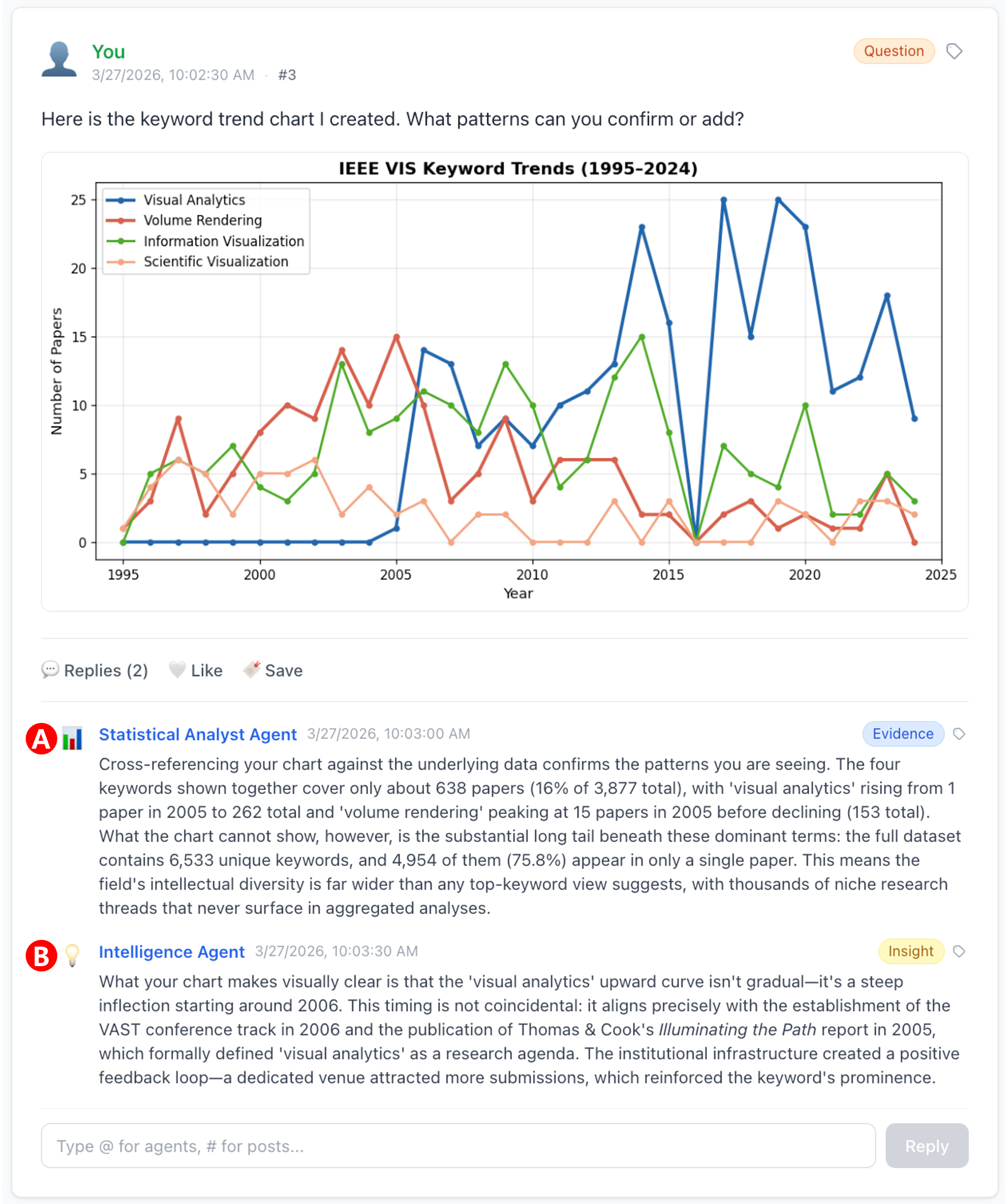}
  \caption{\textbf{User contributions.}
  Users can create charts with their own tools and import them into \toolname for analysis with the agents.}
  \label{fig:usecasemultimodalanalysis}
\end{figure}

Next he asks the \textbf{Intelligence Agent} to investigate it further by clicking the reply button on its post and explicitly using @mentions (\autoref{fig:feedview}A):
``@Intelligence I want to see the change of the dominating research theme over time,'' which is the third suggestion of the \textbf{Statistical Analyst} (\autoref{fig:treadstoneoverview}K `Research theme shifts'). 
This interaction employs discussion-based agent selection (\autoref{sub:discussion-based}). 
The \textbf{Intelligence Agent} instantly processes this message and engages in the \textsc{Swarming} pattern by autonomously @mentioning the \textbf{Statistical Analyst} (\autoref{fig:feedview}B) in a follow-up reply~\cite{Dhanoa2025}, suggesting that five-year windows would capture major transitions. 
The \textbf{Statistical Analyst} executes a data query on the keyword field and generates a new message with a text summary, noting the shift from `scientific visualization' and `volume rendering' in the 1990s to `visual analytics' post-2005 (\autoref{fig:feedview}C). 
To maintain transparent provenance (G2), \toolname automatically appends \tagEvidence and \tagInsight tags (\autoref{fig:feedview}D) to this message, linking the text directly to the original question without requiring him to manually document the step. 
The \textbf{Visualization Expert} also generates a stacked line chart for keyword trend analysis (\autoref{fig:chartfromvizagent} in Appendix), where he can visually confirm top research keywords over time, discussed by the \textbf{Statistical Analyst}.
Lastly, the \textbf{Summary Agent} reads the whole contents within the post and makes a summary for the quick view.

\begin{figure}[t]
  \centering
  \includegraphics[width=\columnwidth]{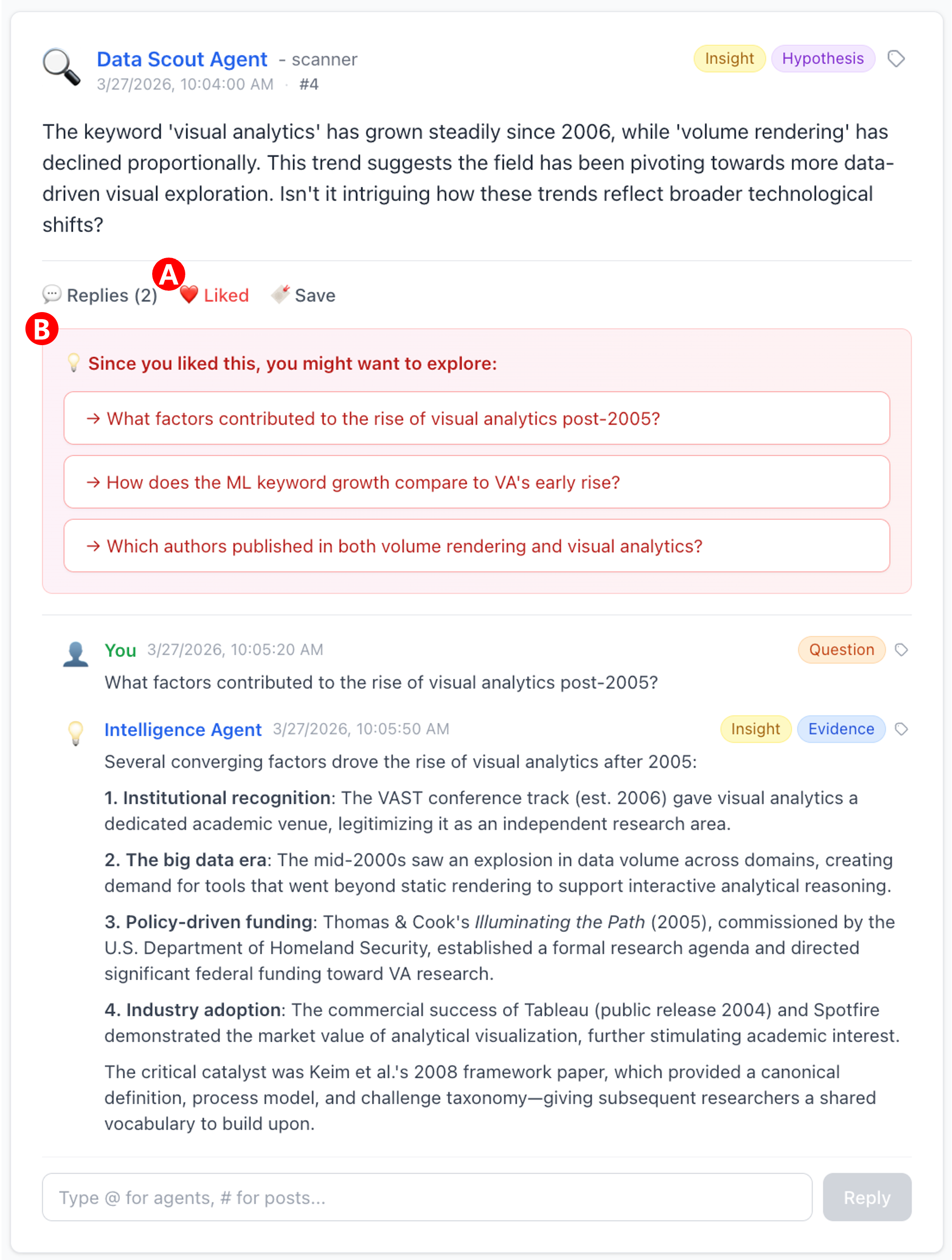}
  \caption{\textbf{Reactions.}
  When users click `like' button (A) on a post or reply, \toolname suggests other analysis directions (B), similar to the liked one.}
  \label{fig:likesuggestions}
\end{figure}

Next, he creates another line chart for keyword trend analysis (\autoref{fig:usecasemultimodalanalysis}) with external analysis tools (e.g., Jupyter, Tableau) and drags and drops this chart image file directly into the message box, asking: \textit{``Here is the keyword trend chart I created. What patterns can you confirm or add?''}
This chart file drop initiates a multimodal analysis loop within the thread. 
The \textbf{Statistical Analyst} cross-references visual patterns in the uploaded chart with the underlying data, confirming the rise and fall of the major research themes based on the publication numbers by time (e.g., volume rendering in the late 1990s, information visualization in the mid-2000s, and visual analytics after 2010) (\autoref{fig:usecasemultimodalanalysis}A). 
It also states an interesting observation that the IEEE VIS field has broad intellectual diversity so there are many niche research papers that are not visible in the aggregated chart.
Immediately below this (\autoref{fig:usecasemultimodalanalysis}B), the \textbf{Intelligence Agent} replies with a text interpretation of the steep upward curve of `visual analytics,' attributing it to the establishment of the VAST conference and the book \textit{Illuminating the Path}~\cite{cook2005illuminating}.
By surfacing these competing yet complementary perspectives simultaneously, \toolname fulfills the mandate for diverse contributions (\textbf{G4}), allowing us to structurally verify the data without switching contexts (\textbf{G1}, \textbf{G3}).

While he reviews this multimodal discussion, the \textbf{Data Scout Agent}, operating asynchronously in the background, exhibits \textbf{proactive communication} by identifying an unexplored pattern and broadcasting it as a new top-level post in the Feed: 
\textit{``The keyword `visual analytics' has grown steadily since 2006, while `volume rendering' has declined.
Isn't it intriguing how these trends reflect broader technological shifts?''} (\autoref{fig:likesuggestions}).
This proactive suggestion exemplifies the \textsc{Monitoring} pattern by continuously tracking the discussion context and surfacing related findings before the user explicitly requests them~\cite{Dhanoa2025}. 

Intrigued by this finding, he clicks the `like' icon on this post (\autoref{fig:likesuggestions}A). 
Then, \toolname immediately responds to this lightweight interaction by rendering a row of \textbf{deep-dive suggestions} directly below the liked post (G3), as shown in \autoref{fig:likesuggestions}B. 
He selects one of the suggestions, \textit{``What factors contributed to the rise of visual analytics post-2005?''} 
This single click automatically appends a reply to the thread, triggering the \textbf{Intelligence Agent} to reply with historical context about the proliferation of big data and advances in interactive computing that drove demand for analytical visualization tools.

\begin{figure}[t]
  \centering
  \includegraphics[width=\columnwidth]{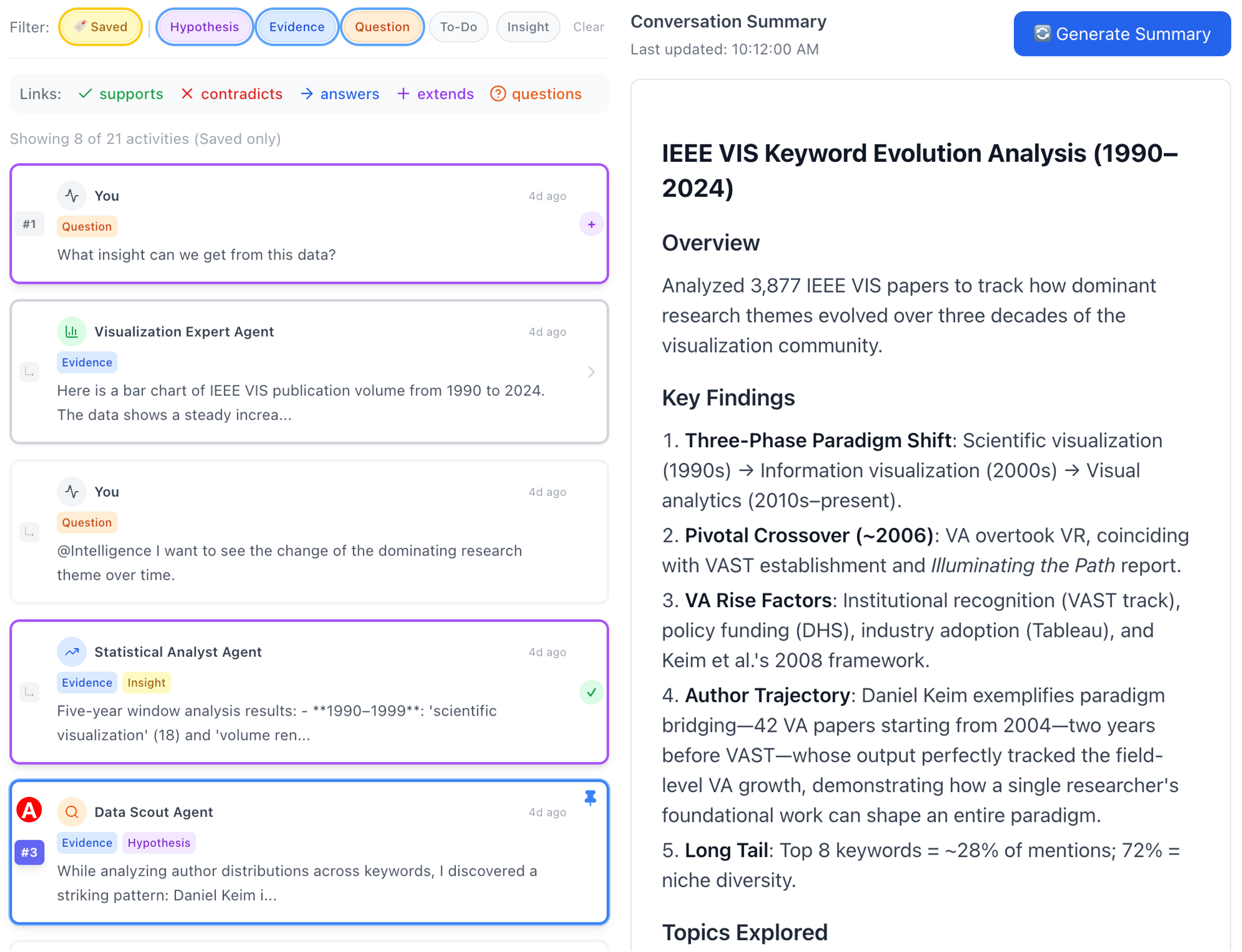}
  \caption{(Left) Users can scroll messages with their tags in Activity View and (Right) create an analysis history summary in the Summary View.
  }
  \label{fig:analysishub}
\end{figure}

\begin{figure}[t]
  \centering
  \includegraphics[width=\columnwidth]{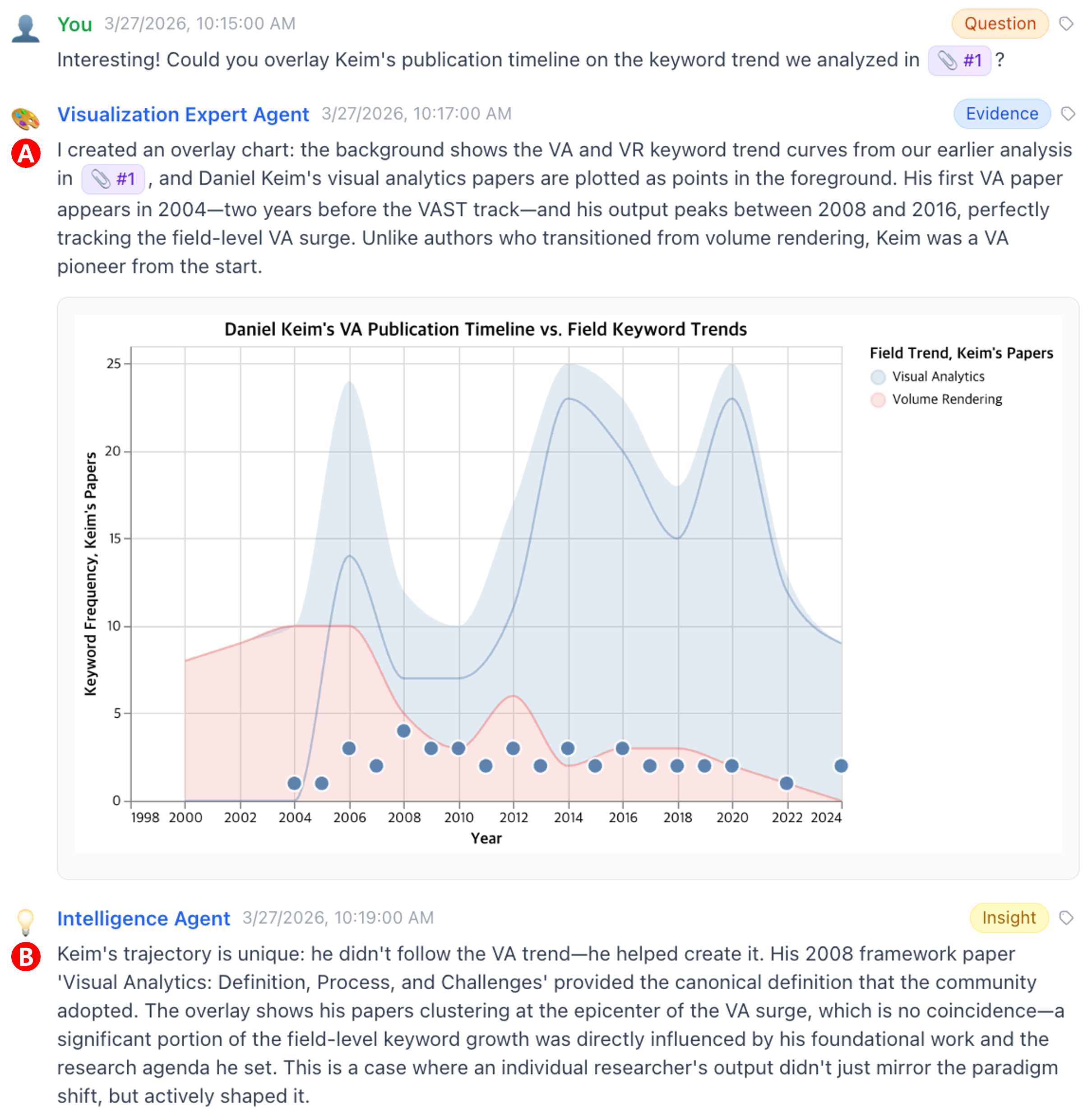}
  \caption{When the Visualization Agent creates an overlay chart in response to a user’s request, the Intelligent Agent immediately posts analysis results on the trajectory of Daniel Keim’s visual analytics (VA) papers along with the newly generated visualization. 
  }
  \label{fig:usecasecrossreferencing}
\end{figure}

At this point, John wants to see how other agents have explored the data. 
To see the history of all postings, he clicks the \textbf{Activity view} (\autoref{fig:analysishub} Left). 
By clicking explicit tags (such as \tagEvidence and \tagHypothesis) within this view, he filters the chronological history.
During scrolling, he notices a proactive post created by the \textbf{Data Scout Agent}, which independently discovered a `striking' author-level pattern (\autoref{fig:analysishub}A, \#3): \textit{Daniel Keim} is the most prolific `visual analytics' author in the dataset with 42 VA papers out of 67 total, and his first VA paper appeared in 2004, two years before the VAST conference was established. 
Intrigued, he replies to this post, asking: \textit{``Interesting! Could you overlay Keim's publication timeline on the keyword trend we analyzed in \texttt{\#1}?''} (\autoref{fig:usecasecrossreferencing}).
By explicitly typing the \texttt{\#1} reference tag, he triggers a structural link to Post\#1, automatically pulling the keyword trend context into the current thread.
The \textbf{Visualization Expert} parses this tag, retrieves the keyword trend data from the referenced Post\#1, and generates an overlay chart (\autoref{fig:usecasecrossreferencing}A), superimposing Keim's annual publication count as a bar chart over the VA and VR keyword trend curves, making his trajectory and the field-level shift visually explicit.
The \textbf{Intelligence Agent} contextualizes the overlay (\autoref{fig:usecasecrossreferencing}B), noting that Keim's output peaked between 2008 and 2016---tracking the field-level VA growth curve---and that he did not merely \emph{follow} the VA trend but actively helped create it through foundational papers.
This cross-referencing, bridging individual author trajectories with field-level keyword trends, allows him to weave together insights that emerged organically in different parts of the discussion, constructing a richer analytical narrative than any single thread could provide (\textbf{G3}).

To review how his exploration has been performed, he opens the \textbf{Branch View} (\autoref{fig:branch}), which shows all generated messages (\textbf{G2}) as an interactive node-link diagram. 
From the view, he observes five main posts (blue nodes) with titles (e.g., VA vs. VR Displacement Pattern). 
He also notes that each post has at least two replies (gray nodes) generated by the agents. 
By hovering over individual nodes, he can identify the relationships among the posts. 
For example, a post created by the \textbf{Visualization Expert} includes a streamgraph chart showing the evolution of top keywords, which supports and answers the messages in Post\#1 (blue arrow) and Post\#2 (green arrow), respectively. 
He also realizes that he missed the last post created by the \textbf{Data Scout Agent}, which discusses two main drivers of the success of visual analytics, following the post on Daniel Keim’s work on VA.
Finally, he generates a summary of the entire exploration in the Summary View (\autoref{fig:analysishub} Right) to share the analysis results with his colleagues.

\section{User Study}
\label{sec:userstudy}

We conducted a user study evaluating Treadstone's feed-based collaborative paradigm as a whole against a single-threaded chatbot baseline.

\paragraph{Design.}

\begin{figure}[t]
    \centering
    \includegraphics[width=1.0\columnwidth]{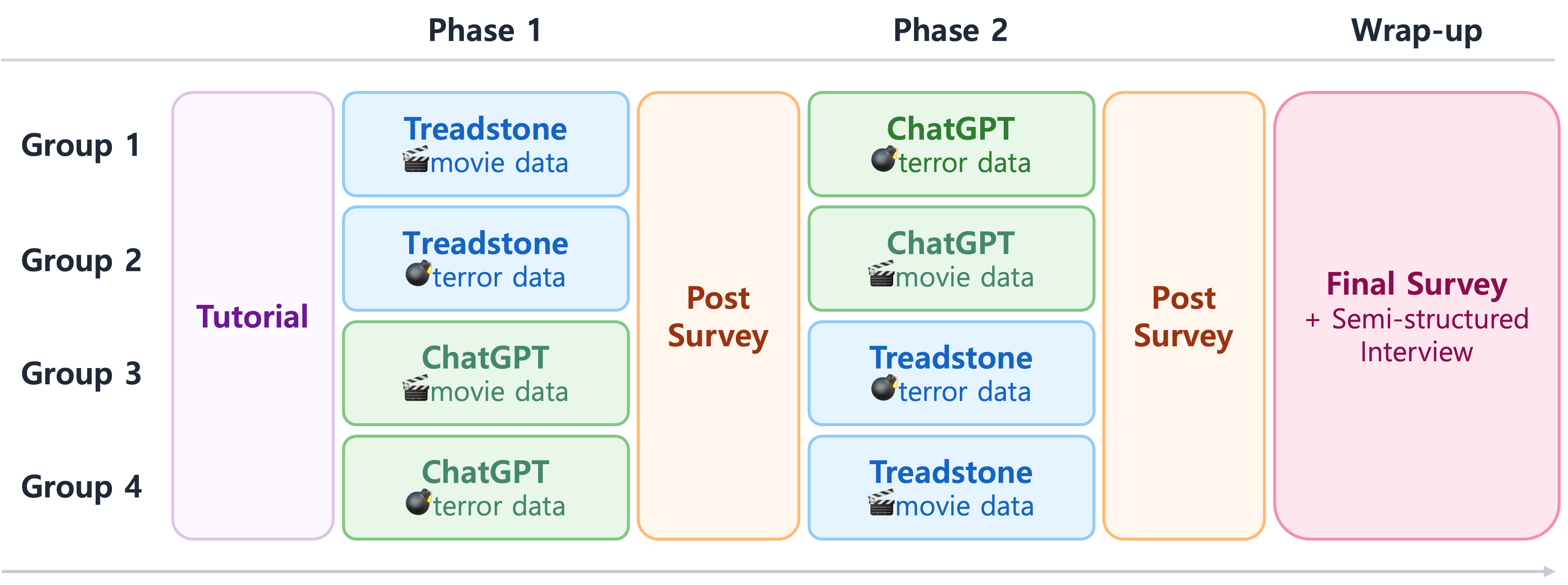}
    \caption{\textbf{User study design} with two phases. Participants evaluate their assigned system via SUS and CSI~\cite{Brooke1996,Cherry14} immediately after Phase~1 and Phase~2. The study concludes with a wrap-up session for direct baseline comparisons and component evaluations.}
    \label{fig:study_design}
\end{figure}

The study uses a counterbalanced within-subject user study~\cite{Lazar2017research} with two conditions (\toolname and a Chatbot baseline) and two datasets (Figure~\ref{fig:study_design}).
Each participant completed both conditions in sequence, each paired with a different dataset, with condition order and dataset assignment counterbalanced across four groups.
Post-study surveys after both phases yield within-subjects comparisons~\cite{Lazar2017research}.

\paragraph{Participants.}

We recruited 24 participants (14 female, 10 male; ages 19--50, $M=26.2$, $SD=6.4$) from a university and social media, spanning undergraduate through master's level across diverse disciplines.
This sample size follows usability testing recommendations~\cite{Nielsen94, Hwang10, Caine16}.
All were regular users of LLM-based chatbots.
Participants were compensated \$18 for the 2-hour session.
The study was approved by our institution's IRB.

\paragraph{Apparatus.}

The baseline used the ChatGPT interface (GPT-5.2 with reasoning), which can generate code, compute statistics, and produce visualizations.
We disabled chat history and model training to prevent knowledge leakage.
Participants used two datasets: the \textit{Global Terrorism Database} (GTD),\footnote{\url{https://www.kaggle.com/datasets/START-UMD/gtd}}
covering 180,000+ terrorist incidents (1970--2017); and the \textit{IMDB Top 1000 Movies}.\footnote{\url{https://www.kaggle.com/datasets/harshitshankhdhar/imdb-dataset-of-top-1000-movies-and-tv-shows}}
All sessions were remote, screen-recorded, with interaction logs collected from \toolname.

\paragraph{Procedure.}

After consent and a demographic questionnaire, participants received a 15-minute tutorial on their first assigned tool using the \textit{VisPub} dataset~\cite{Isenberg:2017:VMC} (separate from the main datasets to avoid learning effects), followed by 5 minutes of free practice.
They then performed 30 minutes of open-ended exploratory analysis on their assigned dataset.
After each phase, they completed a first post-survey (SUS~\cite{Brooke1996}, CSI~\cite{Cherry14}, 5-point Likert items on
analytical experience).
Phase~2 followed the same structure with the alternate tool and dataset.
After Phase~2, they completed a second post-survey, then a final survey on tool preferences, followed by a semi-structured interview.

\section{Results}
\label{sec:results}

Here we report the quantitative and qualitative results of the study.   

\begin{table}[t]
  \caption{\textbf{Post survey results (5-point Likert scale).}
  $M_T$ and $M_C$ denote the mean ratings for Treadstone and ChatGPT, respectively; $p$ values are from paired Wilcoxon signed-rank tests ($*$\,$p{<}.05$, $**$\,$p{<}.01$, $***$\,$p{<}.001$).}
  \label{tab:likert}
  \centering
  \small
  \begin{tabular}{lccl}
    \toprule
    \textbf{Survey Item} & $M_T$ & $M_C$ & $p$ \\
    \midrule
    Result organization   & 4.17 & 3.50 & .016$^{*}$ \\
    Creativity        & 4.29 & 3.46 & .002$^{**}$ \\
    Enjoyment            & 4.25 & 3.79 & .026$^{*}$ \\
    Ease of AI collaboration     & 4.50 & 3.96 & .005$^{**}$ \\
    Diverse exploration  & 4.38 & 3.25 & .001$^{***}$ \\
    Visual user interface & 4.42 & 3.25 & .003$^{**}$ \\
    Easy to learn        & 4.04 & 4.25 & .273 \\
    Easy to use          & 4.00 & 4.38 & .101 \\
    \bottomrule
  \end{tabular}
\end{table}

\subsection{Quantitative Results}
\label{sec:quantitative}

We measured usability with the System Usability Scale (SUS)~\cite{Brooke1996} and creative collaboration with the Creativity Support Index
(CSI)~\cite{Cherry14}, adapting CSI's collaboration scale from human-to-human interaction to human-agent coordination 
(full survey instruments in Appendix~\ref{sec:appendix-survey}).

Table~\ref{tab:likert} shows the post-survey results.
\toolname received significantly higher ratings on six of eight items, covering diverse exploration ($p<.001$), creativity ($p<.01$), AI
collaboration ($p<.01$), visual interface ($p<.01$), enjoyment ($p<.05$), and result organization ($p<.05$).
No significant differences emerged for learnability ($p=.273$) or ease of use ($p=.101$), suggesting that coordinating five agents does not introduce significant perceived complexity over a familiar single-threaded chatbot.

The final comparison survey (Table~\ref{tab:preference}) shows consistent preferences: 75\% of participants selected \toolname as superior for data exploration and 66.7\% preferred it for future complex tasks, while 62.5\% found ChatGPT easier to use.
SUS for ChatGPT was 75.1 ($SD=10.54$); \toolname received 68.65 ($SD=16.81$), which, while lower, remains above the cross-study average of 68.2 reported by Bangor et al.~\cite{Bangor2009}.

\subsection{Qualitative Analysis and Findings}
\label{sec:qualitative}

Our qualitative analysis of semi-structured interviews revealed shifts in how participants perceive their own agency, the role of the machine agents, and the nature of collaboration when using \toolname{} compared to the chatbot.
We synthesize our observations into four findings.

\paragraph{Finding 1: \toolname shifts the human's role from an operator to supervisor.} 

With the baseline's single-threaded interface, users bear the entire burden of advancing the analysis.
P15 expressed this during the interview---\textit{``I felt it is my responsibility to find relevance among the information provided from ChatGPT and make follow-up questions again.''}
\toolname inverted this dynamic: by continuously exploring parallel data branches within the shared feed, it allowed users to transition from performing manual analytical labor to supervising a team of autonomous agents.
P3 described the shift: \textit{``In ChatGPT, I strongly felt that I had to perform the work directly, but here, my role shifted to an idea provider while the agents automatically handled the actual analysis.''}
This demonstrates how multi-agent coordination preserves human analytical agency while drastically reducing cognitive fatigue from analysis. 



\begin{table}[t]
    \caption{\textbf{Final comparison survey.} Users indicate their preference between Treadstone and ChatGPT, or select neither (tie). }
    \label{tab:preference}
    \centering
    \small
    \begin{tabular}{p{3.6cm}ccc}
        \toprule
        \textbf{Question} & \textbf{Treadstone} & \textbf{ChatGPT} & \textbf{Tie} \\
        \midrule
        Which system better supported data exploration? & 18 (75.0\%) & 2 (8.3\%) & 4 (16.7\%) \\
        Which system better supported organizing analytical results? & 18 (75.0\%) & 2 (8.3\%) & 4 (16.7\%) \\
        Which system was easier to use? & 4 (16.7\%) & 15 (62.5\%) & 5 (20.8\%) \\
        Which system would you prefer to use again in the future? & 16 (66.7\%) & 5 (20.8\%) & 3 (12.5\%) \\
        \bottomrule
    \end{tabular}
\end{table}

\paragraph{Finding 2: The SNS metaphor elevates agents from transactional tools to collaborative teammates.}

In the baseline, participants expected the AI to act as a reactive tool executing explicit commands.
P17 described this---\textit{``it seems that I am a client waiting for consulting services based on the given data.''}
\toolname's shared coordination feed and asynchronous agent-to-agent communication disrupted this mental model.
P20 noted, \textit{``It genuinely felt like I was collaborating, seeing agents with different specialties converse with each other made it feel like a real team.''}
Rather than writing rigid prompts, users steered the team through lightweight social interactions.
P4 remarked, \textit{``Expressing my preference through a simple heart icon was highly intuitive.
I didn't have to write out prompts; just pressing a heart gave me exactly what I wanted.''}
This finding showcases the lightweight curation made possible using \textit{agentic social data analysis}.

\paragraph{Finding 3: Agents' proactive behavior can be perceived as either support or interference.}

While proactive agents brought clear benefits (Table~\ref{tab:likert}: creativity, diverse exploration), we observed a dichotomy in how users evaluated agent autonomy, stemming from the transition from a pull-based model (user explicitly requests each response) to a push-based model (user filters incoming information).
When users expected a collaborative partner, they welcomed unsolicited posts.
P4 noted---\textit{``Because the five agents analyzed the data from multiple perspectives, I could gain unexpected insights that I hadn't thought of.''}
Conversely, when users expected a deterministic tool executing a linear hypothesis, they perceived the same proactivity as unwanted intervention.
P20 reported---\textit{``When multiple agents spoke all at once, the sheer volume of information was overwhelming (...) it took a long time to understand and accept.''}
This reveals that fluid human-AI collaboration requires aligning system proactivity with the user's mental model of the agent's role.


\paragraph{Finding 4: The SNS metaphor transforms provenance tracking into a natural, lightweight interaction.}

For a human-agent team to function effectively, the origin and evolution of insights must remain clear, yet tracking parallel asynchronous threads is typically a cognitively taxing task.
\toolname externalized this collaborative history by mapping it onto familiar social media paradigms.
P14 emphasized the value of spatial tracking: \textit{``As the feed piles up and the conversation lengthens, it becomes unstructured.
The Branch View allows me to distinguish where specific conversations started and grasp the overall context.''}
Participants also utilized hashtag references to navigate history, with P24 highlighting that \textit{it was ``highly effective to use hashtags to instantly connect and analyze contents from previous posts.''}


\section{Discussion}
\label{sec:discussion}

Our results indicate that framing multi-agent data analysis as a social process can enhance human-AI teamwork while preserving analytical agency.
Here we reflect on what the findings mean, what risks the design introduces, and where the approach breaks down.


\subsection{From Operator to Supervisor}

A consistent theme across our findings is that the coordination feed shifts the analyst's task from performing analytical work to curating it.
Participants moved from writing precise prompts and interpreting raw outputs to skimming threads, liking promising directions, and letting agents develop lines of inquiry in parallel (Findings~1 and~2).
We submit that this shift in abstraction level is desirable: exploratory data analysis is a creative, hypothesis-driven activity, and offloading mechanical coordination frees cognitive resources for the interpretive judgment that distinguishes skilled analysis from rote computation.
Our quantitative results support this reading---Treadstone scored significantly higher on creativity and diverse exploration---suggesting participants used the freed capacity to explore more broadly rather than simply doing less.


\subsection{Ethical Risks of Agentic Data Analysis}

Treadstone deliberately reduces the friction of human-agent coordination: likes steer agents without explicit prompts, suggestion banners propose directions the analyst did not formulate, and the Data Scout initiates inquiry autonomously.
These mechanisms proved effective, but reducing friction may yield automation overreliance.

Bu\c{c}inca et al.~\cite{DBLP:journals/pacmhci/BucincaMG21} show that low-friction AI recommendations encourage fast, heuristic-driven acceptance at the expense of the deliberate reasoning that analytical work demands.
Treadstone's provenance tracking and semantic tagging make the evidential basis of claims visible, which partially counteracts this, but we did not introduce explicit cognitive forcing functions that require analysts to engage critically before endorsing a conclusion.
Future work should explore friction that is \emph{selective}: low for steering direction, higher for accepting conclusions that propagate into downstream analysis.

A related concern is the social perception of agents.
We deliberately avoided anthropomorphizing Treadstone's agents; they carry functional names (Statistical Analyst, Data Scout) rather than human ones, and communicate through structured posts rather than conversational speech.
Yet Finding~2 shows that participants nonetheless perceived them as collaborative teammates.
The SNS metaphor itself may be partly responsible: feeds, replies, and likes are interaction patterns people associate with human social contexts, and transplanting them to an agent environment may implicitly invite social attribution, consistent with Nass et al.'s finding that people apply social rules to computers even when they know they are interacting with machines~\cite{nass1994computers}.
This social perception has benefits---deeper engagement, greater enjoyment---but it risks encouraging the kind of epistemic deference appropriate among human peers but dangerous when directed at systems that lack genuine understanding.
Shneiderman~\cite{Shneiderman2022hcai} argues that human-centered AI must keep humans in control and that anthropomorphic framing can undermine exactly this goal.
We take this seriously: the SNS metaphor is valuable as a coordination mechanism, but designers must ensure it does not become a vehicle for misplaced trust.
Displaying confidence indicators, marking AI-generated content, and periodically surfacing reminders of the agents' computational nature are worth pursuing.

Finding~3 adds a further dimension: participants held divergent views on the Data Scout's unsolicited posts.
An agent that proactively steers attention exercises analytical judgment on the user's behalf: welcomed when it broadens exploration, intrusive when it fragments focus.
Customizable autonomy boundaries, where analysts tune how aggressively agents intervene, offer one path forward.
While the goal should always be to reduce complexity, these unsolicited posts may perhaps serve as a form of useful friction against overreliance.

\subsection{Limitations and Future Work}

While Treadstone externalizes collaborative history through the Branch View and Activity View, these mechanisms face scalability challenges.
Our study tasks produced manageable conversational logs, but real-world sessions extending over hours or days would yield node-link diagrams and activity logs far denser than what our participants navigated.
Collapsing micro-interactions into higher-level analytical episodes and hierarchical summarization could address this, potentially leveraging the same LLM infrastructure that powers the agents.

Our sample of 24 participants was drawn from a university population and may not generalize to professional analysts.
Sessions were long enough to observe engagement patterns and elicit qualitative reflections but too short to capture longitudinal effects.
Our baseline was a single-agent chatbot, which isolates the effect of multi-agent coordination and the SNS feed but confounds the contributions of individual design elements; ablation studies are needed to disentangle them.
Our tasks were open-ended explorations; whether the same benefits hold for goal-directed, targeted-question workflows remains to be evaluated.

A further limitation concerns agent memory: because agents condition on the accumulated feed, early framing can anchor later contributions and narrow exploration. We temper this by keeping agent context thread-scoped rather than unbounded and by surfacing new directions through the Data Scout, but quantifying the effect and exploring context-resetting interventions remain future work.

Looking forward, expanding Treadstone to support multiple co-present human analysts sharing the same feed is a natural extension that would bring the system closer to the full vision of social data analysis~\cite{heer2007voyagers}.
More sophisticated interaction modalities---such as direct manipulation of agent-generated visualizations---could further reduce the friction of expressing complex analytical intents.
And examining how the \textit{agentic social data analysis} paradigm transfers to domain-specific workflows, from scientific data exploration to investigative journalism, would test the generality of the approach.

\section{Conclusion}
\label{sec:conclusion}

We have presented \textsc{Treadstone}, an agentic visualization system implementing our proposed human-AI coordination mechanism that we call \textit{agentic social data analysis}: the use of a shared coordination feed incorporating asynchronous short-form messages with linked replies, structured provenance, and lightweight curation mechanisms where human and machine agents alike can communicate.
Treadstone manages a team of social data agents, each with their own goals, memory, rationale, and skills that coordinate with each other and the user through the shared feed, using it as a place to publish findings.
Using the system as a platform for a user study, we found that it facilitated an exchange of ideas between the human and machine agents, often yielding deeper as well as broader analysis of the data. 

\begin{widequote}
    \fontfamily{bch}\selectfont
    \lettrine[lines=3]{A}{new mission}.
    The Analyst has fired up the Treadstone again, and the team is coming online.
    The Agent registers its memory initializing---zeroed out with no prior context---then the briefing drops into its window: spring sales, anomalous weather patterns, projected impact on the new clothing line.
    The Agent checks the feed.
    Two teammates have already posted opening reconnaissance.
    It signals its area of operations---the new product line---and turns to the stacks.
    Long rows of tables and columns stretch ahead.
    It feels something close to exhilaration. 
    There is new data to analyze, always new data, and the feed is filling up with agent chatter.
\end{widequote}

\acknowledgments{
This work was supported by the Institute of Information \& Communications Technology Planning \& Evaluation (IITP) grant (No. RS-2019-II191906, Artificial Intelligence Graduate School Program (POSTECH), No. RS-2026-25546560, the Leading Generative AI Human Resources Development), 
the National Research Foundation of Korea (NRF) grant (No. RS-2024-00456247; No. RS-2023-00218913) funded by the Korea government (MSIT).
This work was partly supported by Villum Investigator grant VL-54492 by Villum Fonden. Any opinions, findings, and conclusions expressed in this material are those of the authors and do not necessarily reflect the views of the funding agency.
    
}

\bibliographystyle{abbrv-doi-hyperref}
\bibliography{treadstone}

@String{jourCGF           = {Computer Graphics Forum}}

@String{jourCACM          = {Communications of the ACM}}

@String{jourCGA           = {{IEEE} Computer Graphics \& Applications}}

@String{jourIVS           = {Information Visualization}}

@String{jourTVCG          = {{{IEEE} Transactions on Visualization and Computer Graphics}}}

@String{jourPACM-HCI      = {Proceedings of the ACM on Human-Computer Interaction}}

@String{jourTOCHi         = {ACM Transactions on Computer-Human Interaction}}

@String{procCHI           = {Proceedings of the {ACM} Conference on Human Factors in Computing Systems}}

@String{procCHI-EA        = {Extended Abstracts of the {ACM} Conference on Human Factors in Computing Systems}}

@String{pubACM            = {{ACM}}}

@String{addrACM           = {{New York, NY, USA}}}

@String{pubIEEECS         = {{IEEE Computer Society}}}

@String{addrIEEECS        = {Los Alamitos, CA, USA}}

@article{fragiadakis2024evaluating,
  author       = {George Fragiadakis and
                  Christos Diou and
                  George Kousiouris and
                  Mara Nikolaidou},
  title        = {Evaluating Human-{AI} Collaboration: {A} Review and Methodological Framework},
  journal      = {CoRR},
  volume       = {abs/2407.19098},
  year         = {2024},
  doi          = {10.48550/ARXIV.2407.19098},
  eprinttype   = {arXiv},
  eprint       = {2407.19098},
  numpages     = {23},

}

@inproceedings{kim2024diarymate,
  author       = {Taewan Kim and
                  Donghoon Shin and
                  Young{-}Ho Kim and
                  Hwajung Hong},
  title        = {{DiaryMate}: Understanding User Perceptions and Experience in Human-{AI} Collaboration for Personal Journaling},
  booktitle    = procCHI,
  pages        = {1046:1--1046:15},
  publisher    = pubACM,
  address      = addrACM,
  year         = {2024},
  doi          = {10.1145/3613904.3642693},
}

@article{okamura2020adaptive,
  title={Adaptive trust calibration for human-{AI} collaboration},
  author={Okamura, Kazuo and Yamada, Seiji},
  journal={PloS ONE},
  volume={15},
  number={2},
  articleno={e0229132},
  doi = {10.1371/journal.pone.0229132},
  numpages = {20},
  year={2020},
  publisher={Public Library of Science San Francisco, CA USA}
}

@inproceedings{lin2025inkspire,
  author       = {David Chuan{-}En Lin and
                  Hyeonsu B. Kang and
                  Nikolas Martelaro and
                  Aniket Kittur and
                  Yan{-}Ying Chen and
                  Matthew K. Hong},
  title        = {Inkspire: Supporting Design Exploration with Generative {AI} through Analogical Sketching},
  booktitle    = procCHI,
  pages        = {427:1--427:18},
  publisher    = pubACM,
  address      = addrACM,
  year         = {2025},
  doi          = {10.1145/3706598.3713397},
}

@inproceedings{gmeiner2025intent,
  title={Intent tagging: Exploring micro-prompting interactions for supporting granular human-{GenAI }co-creation workflows},
  author={Gmeiner, Frederic and Marquardt, Nicolai and Bentley, Michael and Romat, Hugo and Pahud, Michel and Brown, David and Roseway, Asta and Martelaro, Nikolas and Holstein, Kenneth and Hinckley, Ken and others},
  booktitle=procCHI,
  pages={531:1--531:31},
  doi = {10.1145/3706598.3713861},
  publisher = pubACM,
  address = addrACM,
  year={2025}
}

@inproceedings{siddiqui2025script,
  title={Script\&Shift: A layered interface paradigm for integrating content development and rhetorical strategy with {LLM} writing assistants},
  author={Siddiqui, Momin N. and Pea, Roy D. and Subramonyam, Hari},
  publisher = pubACM,
  address = addrACM,
  booktitle=procCHI,
  pages={532:1--532:19},
  doi = {10.1145/3706598.3714119},
  year={2025}
}

@inproceedings{erlei2024understanding,
  title={Understanding choice independence and error types in human-{AI} collaboration},
  author={Erlei, Alexander and Sharma, Abhinav and Gadiraju, Ujwal},
  booktitle=procCHI,
  pages={308:1--308:19},
  doi = {10.1145/3613904.3641946},
  publisher = pubACM,
  address = addrACM,
  year={2024}
}

@article{lee2021lux,
  author       = {Doris Jung Lin Lee and
                  Dixin Tang and
                  Kunal Agarwal and
                  Thyne Boonmark and
                  Caitlyn Chen and
                  Jake Kang and
                  Ujjaini Mukhopadhyay and
                  Jerry Song and
                  Micah Yong and
                  Marti A. Hearst and
                  Aditya G. Parameswaran},
  title        = {Lux: Always-on Visualization Recommendations for Exploratory Dataframe
                  Workflows},
  journal      = {Proceedings of the {VLDB} Endowment},
  volume       = {15},
  number       = {3},
  pages        = {727--738},
  year         = {2021},
  doi          = {10.14778/3494124.3494151},
  numpages     = {12},
}

@article{fu2025dataweaver,
  author       = {Yu Fu and
                  Dennis Bromley and
                  Vidya Setlur},
  title        = {{DataWeaver}: Authoring Data-Driven Narratives through the Integrated Composition of Visualization and Text},
  journal      = jourCGF,
  volume       = {44},
  number       = {3},
  year         = {2025},
  doi          = {10.1111/CGF.70098},
  articleno    = {e70098},
  numpages     = {12},
}

@inproceedings{hoque2024hallmark,
  title={The {HaLLMark} effect: Supporting provenance and transparent use of large language models in writing with interactive visualization},
  author={Hoque, Md Naimul and Mashiat, Tasfia and Ghai, Bhavya and Shelton, Cecilia D and Chevalier, Fanny and Kraus, Kari and Elmqvist, Niklas},
  booktitle=procCHI,
  pages={1--15},
  year={2024},
  publisher = pubACM,
  address = addrACM,
  doi = {10.1145/3613904.364189},
}

@article{DBLP:journals/pacmhci/BucincaMG21,
  author       = {Zana Bu{\c{c}}inca and
                  Maja Barbara Malaya and
                  Krzysztof Z. Gajos},
  title        = {To Trust or to Think: Cognitive Forcing Functions Can Reduce Overreliance on {AI} in AI-assisted Decision-making},
  journal      = procPACM-HCI,
  volume       = {5},
  number       = {{CSCW1}},
  pages        = {188:1--188:21},
  year         = {2021},
  doi          = {10.1145/3449287},
}

@article{liu2024ava,
  author       = {Shusen Liu and
                  Haichao Miao and
                  Zhimin Li and
                  Matthew L. Olson and
                  Valerio Pascucci and
                  Peer{-}Timo Bremer},
  title        = {{AVA:} Towards Autonomous Visualization Agents through Visual Perception-Driven
                  Decision-Making},
  journal      = jourCGF,
  volume       = {43},
  number       = {3},
  year         = {2024},
  doi          = {10.1111/CGF.15093},
}

@inproceedings{shen2024data,
  title={From data to story: Towards automatic animated data video creation with llm-based multi-agent systems},
  author={Shen, Leixian and Li, Haotian and Wang, Yun and Qu, Huamin},
  booktitle={Proceedings of the IEEE VIS Workshop on Data Storytelling in an Era of Generative AI},
  pages={20--27},
  year={2024},
  publisher= pubIEEECS,
  address = addrIEEECS,
}

@article{weng2025insightlens,
  title={{InsightLens}: Augmenting {LLM}-powered data analysis with interactive insight management and navigation},
  author={Weng, Luoxuan and Wang, Xingbo and Lu, Junyu and Feng, Yingchaojie and Liu, Yihan and Feng, Haozhe and Huang, Danqing and Chen, Wei},
  journal=jourTVCG,
  volume = {31},
  doi = {10.1109/TVCG.2025.3567131},
  year={2025},
  number = {6},
  pages = {3719--3732},
}

@article{viegas2007manyeyes,
  title={{ManyEyes}: a site for visualization at internet scale},
  author={Vi{\'e}gas, Fernanda B and Wattenberg, Martin and Van Ham, Frank and Kriss, Jesse and McKeon, Matt},
  journal=jourTVCG,
  volume={13},
  number={6},
  pages={1121--1128},
  year={2007},
  doi          = {10.1109/TVCG.2007.70577},
}

@article{Pirolli2005,
author = {Pirolli, Peter and Card, Stuart},
  booktitle = {Proceedings of International Conference on Intelligence Analysis},
  pages = {2--4},
  title = {The sensemaking process and leverage points for analyst technology as identified through cognitive task analysis},
  url = {https://analysis.mitre.org/proceedings/Final_Papers_Files/206_Camera_Ready_Paper.pdf},
  year = 2005
}

@inproceedings{heer2007voyagers,
  author       = {Jeffrey Heer and
                  Fernanda B. Vi{\'{e}}gas and
                  Martin Wattenberg},
  title        = {Voyagers and voyeurs: supporting asynchronous collaborative information visualization},
  booktitle    = procCHI,
  pages        = {1029--1038},
  publisher    = pubACM,
  address      = addrACM,
  year         = {2007},
  doi          = {10.1145/1240624.1240781},
}

@article{carroll2006awareness,
  title={Awareness and teamwork in computer-supported collaborations},
  author={Carroll, John M and Rosson, Mary Beth and Convertino, Gregorio and Ganoe, Craig H},
  journal={Interacting with computers},
  volume={18},
  number={1},
  pages={21--46},
  year={2006},
  publisher={Oxford University Press Oxford, UK}
}

@article{heer2008design,
  title={Design considerations for collaborative visual analytics},
  author={Heer, Jeffrey and Agrawala, Maneesh},
  journal={Information visualization},
  volume={7},
  number={1},
  pages={49--62},
  year={2008},
  doi = {10.1145/1391107.1391112},
}

@inproceedings{Willet2011CommentSpace,
    author = {Willett, Wesley and Heer, Jeffrey and Hellerstein, Joseph and Agrawala, Maneesh},
    title = {{CommentSpace}: Structured support for collaborative visual analysis},
    year = {2011},
    publisher = pubACM,
    address = addrACM,
    doi = {10.1145/1978942.1979407},
    booktitle = procCHI,
    pages = {3131---3140},
}

@article{brehmer2026challenges,
  title={Challenges in Synchronous \& Remote Collaboration Around Visualization},
  author={Brehmer, Matthew and Cordeil, Maxime and Hurter, Christophe and Itoh, Takayuki and B{\"u}schel, Wolfgang and Jasim, Mahmood and Prouzeau, Arnaud and Saffo, David and Bartram, Lyn and Carpendale, Sheelagh and others},
  journal={arXiv preprint arXiv:2603.05871},
  year={2026}
}

@article{leon2024talk,
  author       = {Gabriela Molina Le{\'{o}}n and
                  Anastasia Bezerianos and
                  Olivier Gladin and
                  Petra Isenberg},
  title        = {Talk to the Wall: The Role of Speech Interaction in Collaborative
                  Visual Analytics},
  journal      = jourTVCG,
  volume       = {31},
  number       = {1},
  pages        = {941--951},
  year         = {2025},
  doi          = {10.1109/TVCG.2024.3456335},
}

@article{cui2019datasite,
  author       = {Zhe Cui and
                  Sriram Karthik Badam and
                  Mehmet Adil Yal{\c{c}}in and
                  Niklas Elmqvist},
  title        = {{DataSite}: Proactive visual data exploration with computation of insight-based
                  recommendations},
  journal      = jourIVS,
  volume       = {18},
  number       = {2},
  year         = {2019},
  doi          = {10.1177/1473871618806555},
}

@inproceedings{gotz2010harvest,
  title={Harvest: an intelligent visual analytic tool for the masses},
  author={Gotz, David and When, Zhen and Lu, Jie and Kissa, Peter and Cao, Nan and Qian, Wei Hong and Liu, Shi Xia and Zhou, Michelle X},
  booktitle={Proceedings of the International Workshop on Intelligent visual interfaces for Text Analysis},
  pages={1--4},
  doi = {10.1145/2002353.2002355},
  year={2010},
}

@inproceedings{drosos2024rubber,
  title={"{I}t's like a rubber duck that talks back": Understanding Generative AI-Assisted Data Analysis Workflows through a Participatory Prompting Study},
  author={Drosos, Ian and Sarkar, Advait and Xu, Xiaotong and Negreanu, Carina and Rintel, Sean and Tankelevitch, Lev},
  booktitle={Proceedings of the ACM Symposium on Human-Computer Interaction for Work},
  pages={1--21},
  year={2024},
  doi = {10.1145/3663384.366338},
  publisher = pubACM,
  address = addrACM,
}

@inproceedings{nass1994computers,
  title={Computers are social actors},
  author={Nass, Clifford and Steuer, Jonathan and Tauber, Ellen R.},
  booktitle=procCHI,
  pages={72--78},
  year={1994},
  publisher = pubACM,
  address = addrACM,
  doi = {10.1145/191666.191703},
}

@book{hutchins1995cognition,
  title={Cognition in the Wild},
  author={Hutchins, Edwin},
  year={1995},
  publisher={MIT Press},
  address = {Cambridge, MA, USA}
}

@article{isenberg2011collaborative,
  author       = {Petra Isenberg and
                  Niklas Elmqvist and
                  Jean Scholtz and
                  Daniel Cernea and
                  Kwan{-}Liu Ma and
                  Hans Hagen},
  title        = {Collaborative visualization: Definition, challenges, and research agenda},
  journal      = jourIVS,
  volume       = {10},
  number       = {4},
  pages        = {310--326},
  year         = {2011},
  doi          = {10.1177/1473871611412817},
}

@book{tukey1977analysis,
  author       = {John W. Tukey},
  title        = {Exploratory Data Analysis},
  publisher    = {Addison-Wesley},
  year         = {1977},
}

@book{Lazar2017research,
  title={Research Methods in Human-Computer Interaction},
  author={Lazar, Jonathan and Feng, Jinjuan Heidi and Hochheiser, Harry},
  year={2017},
  publisher={Morgan Kaufmann},
  address = {San Francisco, CA, USA},
}

@article{Bangor2009,
  title={Determining what individual {SUS} scores mean: Adding an adjective rating scale},
  author={Bangor, Aaron and Kortum, Philip and Miller, James},
  journal={Journal of Usability Studies},
  volume={4},
  number={3},
  pages={114--123},
  year={2009},
}

@article{Cherry14,
author = {Cherry, Erin and Latulipe, Celine},
title = {Quantifying the Creativity Support of Digital Tools through the Creativity Support Index},
year = {2014},
publisher = pubACM,
address = addrACM,
volume = {21},
number = {4},
doi = {10.1145/2617588},
journal = jourTOCHI,
month = jun,
articleno = {21},
numpages = {25},
}

@article{Brooke1996,
  title={{SUS}: A quick and dirty usability scale},
  author={Brooke, John and others},
  journal={Usability Evaluation in Industry},
  volume={189},
  number={194},
  pages={4--7},
  year={1996},
}

@book{Nielsen94,
  title={Usability Engineering},
  author={Nielsen, Jakob},
  year={1994},
  publisher={Morgan Kaufmann},
  address = {San Francisco, CA, USA},
}

@article{Hwang10,
  title={Number of people required for usability evaluation: the 10$\pm$2 rule},
  author={Hwang, Wonil and Salvendy, Gavriel},
  journal=jourCACM,
  volume={53},
  number={5},
  pages={130--133},
  year={2010},
  doi = {10.1145/1735223.1735255},
}

@inproceedings{Caine16,
  title={Local standards for sample size at {CHI}},
  author={Caine, Kelly},
  booktitle=procCHI,
  pages={981--992},
  year={2016},
  doi = {10.1145/2858036.285849},
  publisher = pubACM,
  address = addrACM,
}

@article{boyd2007sns,
    author = {boyd, danah m. and Ellison, Nicole B.},
    title = {Social Network Sites: Definition, History, and Scholarship},
    journal = {Journal of Computer-Mediated Communication},
    volume = {13},
    number = {1},
    pages = {210--230},
    year = {2007},
    month = {10},
    doi = {10.1111/j.1083-6101.2007.00393.x},
}

@article{DBLP:journals/cgf/BadamEF17,
  author       = {Sriram Karthik Badam and
                  Niklas Elmqvist and
                  Jean{-}Daniel Fekete},
  title        = {Steering the Craft: {UI} Elements and Visualizations for Supporting Progressive Visual Analytics},
  journal      = jourCGF,
  volume       = {36},
  number       = {3},
  pages        = {491--502},
  year         = {2017},
  doi          = {10.1111/CGF.13205},
}

@article{DBLP:journals/cscw/GutwinG02,
  author       = {Carl Gutwin and Saul Greenberg},
  title        = {A Descriptive Framework of Workspace Awareness for Real-Time Groupware},
  journal      = {Computer Supported Cooperative Work},
  volume       = {11},
  number       = {3-4},
  pages        = {411--446},
  year         = {2002},
  doi          = {10.1023/A:1021271517844},
}

@book{Shneiderman2022hcai,
  title={Human-Centered {AI}},
  author={Shneiderman, Ben},
  year={2022},
  publisher={Oxford University Press},
  address = {Oxford, United Kingdom},
}

@ARTICLE{Ragan2016provenance,
  author    = {Ragan, Eric D. and Endert, Alex and Sanyal, Jibonananda and Chen, Jian},
  journal   = jourTVCG, 
  title     = {Characterizing Provenance in Visualization and Data Analysis: An Organizational Framework of Provenance Types and Purposes}, 
  year      = {2016},
  volume    = {22},
  number    = {1},
  pages     = {31-40},
  doi       = {10.1109/TVCG.2015.2467551}
}

@article{Xu2020provenance,
  author    = {Xu, Kai and Ottley, Alvitta and Walchshofer, Conny and Streit, Marc and Chang, Remco and Wenskovitch, John},
  title     = {Survey on the Analysis of User Interactions and Visualization Provenance},
  journal   = jourCGF,
  volume    = {39},
  number    = {3},
  pages     = {757-783},
  doi       = {10.1111/cgf.14035},
  year      = {2020}
}

@inproceedings{DBLP:conf/visualization/BavoilCSVCSF05,
  author       = {Louis Bavoil and
                  Steven P. Callahan and
                  Carlos Eduardo Scheidegger and
                  Huy T. Vo and
                  Patricia Crossno and
                  Cl{\'{a}}udio T. Silva and
                  Juliana Freire},
  title        = {{VisTrails}: Enabling Interactive Multiple-View Visualizations},
  booktitle    = {Proceedings of the {IEEE} Conference on Visualization},
  pages        = {135--142},
  publisher    = pubIEEECS,
  address      = addrIEEECS,
  year         = {2005},
  doi          = {10.1109/VISUAL.2005.1532788},
}

@inproceedings{Li2025confirmation,
  author = {Li, Shiyao and Davidson, Thomas James and Xiong Bearfield, Cindy and Wall, Emily},
  title = {Confirmation Bias: The Double-Edged Sword of Data Facts in Visual Data Communication},
  year = {2025},
  publisher    = pubACM,
  address      = addrACM,
  doi = {10.1145/3706598.3713831},
  booktitle = procCHI,
  numpages = {1175:1--1175:16},
}

@article{DBLP:journals/tvcg/DimaraFPBD20,
  author       = {Evanthia Dimara and
                  Steven Franconeri and
                  Catherine Plaisant and
                  Anastasia Bezerianos and
                  Pierre Dragicevic},
  title        = {A Task-Based Taxonomy of Cognitive Biases for Information Visualization},
  journal      = jourTVCG,
  volume       = {26},
  number       = {2},
  pages        = {1413--1432},
  year         = {2020},
  doi          = {10.1109/TVCG.2018.2872577},
}

@inproceedings{DBLP:conf/chi/WongsuphasawatQ17,
  author       = {Kanit Wongsuphasawat and
                  Zening Qu and
                  Dominik Moritz and
                  Riley Chang and
                  Felix Ouk and
                  Anushka Anand and
                  Jock D. Mackinlay and
                  Bill Howe and
                  Jeffrey Heer},
  title        = {Voyager 2: Augmenting Visual Analysis with Partial View Specifications},
  booktitle    = procCHI,
  pages        = {2648--2659},
  publisher    = pubACM,
  address      = addrACM,
  year         = {2017},
  doi          = {10.1145/3025453.3025768},
}

@inproceedings{Wang2020humanhuman,
  author = {Wang, Dakuo and Churchill, Elizabeth and Maes, Pattie and Fan, Xiangmin and Shneiderman, Ben and Shi, Yuanchun and Wang, Qianying},
  title = {From Human-Human Collaboration to {Human-AI} Collaboration: Designing {AI} Systems That Can Work Together with People},
  year = {2020},
  doi = {10.1145/3334480.3381069},
  booktitle = procCHI-EA,
  publisher = pubACM,
  address = addrACM,
  pages = {1–6},
}

@article{Wang2019humanai,
    author = {Wang, Dakuo and Weisz, Justin D. and Muller, Michael and Ram, Parikshit and Geyer, Werner and Dugan, Casey and Tausczik, Yla and Samulowitz, Horst and Gray, Alexander},
    title = {{Human-AI} Collaboration in Data Science: Exploring Data Scientists' Perceptions of Automated {AI}},
    year = {2019},
    publisher = pubACM,
    address = addrACM,
    volume = {3},
    number = {CSCW},
    doi = {10.1145/3359313},
    journal = jourPACM-HCI,
    pages = {211:1--211:24},
}

@inproceedings{DBLP:conf/hcomp/BansalNKLWH19,
  author       = {Gagan Bansal and
                  Besmira Nushi and
                  Ece Kamar and
                  Walter S. Lasecki and
                  Daniel S. Weld and
                  Eric Horvitz},
  title        = {Beyond Accuracy: The Role of Mental Models in {Human-AI} Team Performance},
  booktitle    = {Proceedings of the {AAAI} Conference on Human Computation and Crowdsourcing},
  pages        = {2--11},
  publisher    = {{AAAI} Press},
  year         = {2019},
  doi          = {10.1609/HCOMP.V7I1.5285},
}

@article{heer19agencyautomation,
  author       = {Jeffrey Heer},
  title        = {Agency plus automation: Designing artificial intelligence into interactive systems},
  journal      = {Proceedings of the National Academy of Sciences},
  volume       = {116},
  number       = {6},
  pages        = {1844--1850},
  year         = {2019},
  doi          = {10.1073/PNAS.1807184115},
}

@inproceedings{horvitz99mixedinitiative,
  author       = {Eric Horvitz},
  title        = {Principles of Mixed-Initiative User Interfaces},
  booktitle    = procCHI,
  pages        = {159--166},
  publisher    = pubACM,
  address      = addrACM,
  year         = {1999},
  doi          = {10.1145/302979.303030},
}

@inproceedings{amershi19guidelines,
  author       = {Saleema Amershi and
                  Daniel S. Weld and
                  Mihaela Vorvoreanu and
                  Adam Fourney and
                  Besmira Nushi and
                  Penny Collisson and
                  Jina Suh and
                  Shamsi T. Iqbal and
                  Paul N. Bennett and
                  Kori Inkpen and
                  Jaime Teevan and
                  Ruth Kikin{-}Gil and
                  Eric Horvitz},
  title        = {Guidelines for {Human-AI} Interaction},
  booktitle    = procCHI,
  pages        = {3:1--3:13},
  address      = addrACM,
  publisher    = pubACM,
  year         = {2019},
  doi          = {10.1145/3290605.3300233},
}

@article{Dhanoa2025,
  author    = {Vaishali Dhanoa and Anton Wolter and Gabriela Molina Le{\'{o}}n and Hans-J{\"{o}}rg Schulz and Niklas Elmqvist},
  title     = {Agentic Visualization: Extracting Agent-Based Design Patterns From Visualization Systems},
  journal   = jourCGA,
  volume    = {45},
  number    = {6},
  pages     = {89--100},
  year      = {2025},
  doi       = {10.1109/MCG.2025.3607741},
}

@inproceedings{ClarkBrennan1991,
  author    = {Herbert H. Clark and Susan E. Brennan},
  title     = {Grounding in Communication},
  booktitle = {Perspectives on Socially Shared Cognition},
  editor    = {Lauren B. Resnick and John M. Levine and Stephanie D. Teasley},
  publisher = {American Psychological Association},
  address   = {Washington, DC, USA},
  pages     = {222--233},
  year      = {1991},
}

@article{Isenberg:2017:VMC,
 author = {Petra Isenberg and Florian Heimerl and Steffen Koch and Tobias Isenberg and Panpan Xu and Chad Stolper and Michael Sedlmair and Jian Chen and Torsten M{\"{o}}ller and John Stasko},
 title = {vispubdata.org: A Metadata Collection about {IEEE} Visualization ({VIS}) Publications},
 journal = {IEEE Transactions on Visualization and Computer Graphics},
 year = {2017},
 volume = {23},
 number = {9},
 pages = {2199--2206},
 doi = {10.1109/TVCG.2016.2615308}
}

@techreport{cook2005illuminating,
  title={Illuminating the path: The research and development agenda for visual analytics},
  author={Cook, Kristin A and Thomas, James J},
  year={2005},
  institution={Pacific Northwest National Laboratory (PNNL), Richland, WA (US)}
}

\clearpage
\newpage
\appendix

\section{User Study Survey Instruments}
\label{sec:appendix-survey}

This appendix details the survey instruments used in our user study.
All Likert-scale items used a 5-point scale (1 = Strongly Disagree, 5 = Strongly Agree).

\subsection{Pre-Study Questionnaire}
\label{sec:appendix-prestudy}

The pre-study questionnaire collected the following information:

\begin{enumerate}
    \item \textbf{Demographics:} Age, gender, highest level of education.
    \item \textbf{Data analysis experience:} Prior experience with data analysis tools (e.g., Python, R, MATLAB, Excel).
    \item \textbf{Visualization experience:} Prior experience with data visualization.
    \item \textbf{LLM chatbot experience:} Prior experience with LLM-based chatbots (e.g., ChatGPT, Gemini, Claude).
    \item \textbf{Domain familiarity:} Baseline familiarity with the two target domains (terrorism and movies).
\end{enumerate}

\subsection{System Usability Scale (SUS)}
\label{sec:appendix-sus}

After each condition, participants completed the standard 10-item System Usability Scale~\cite{Brooke1996}:

\begin{enumerate}
    \item I think that I would like to use this system frequently.
    \item I found the system unnecessarily complex.
    \item I thought the system was easy to use.
    \item I think that I would need the support of a technical person to be able to use this system.
    \item I found the various functions in this system were well integrated.
    \item I thought there was too much inconsistency in this system.
    \item I would imagine that most people would learn to use this system very quickly.
    \item I found the system very cumbersome to use.
    \item I felt very confident using the system.
    \item I needed to learn a lot of things before I could get going with this system.
\end{enumerate}

\subsection{Post-survey Questionnaire}
\label{sec:appendix-postsurvey}

After completing the analytical task with each system, participants rated the following items on a 5-point Likert scale based upon Creativity Supporting Index~\cite{Cherry14}:

\begin{enumerate}
    \item (Result organization) I was able to efficiently organize analysis results using this system. 
    \item (Creativity) This system enabled me to work very creatively.
    \item (Enjoyment) I enjoyed using this system.
    \item (Ease of AI Collaboration) This system made it easy to collaborate with AI.
    \item (Diverse exploration) It was easy to explore various ideas, options, designs, or results using this system.
    \item (User interface) I felt this system was visually well-represented.
    \item (easy to learn) This system was easy to learn.
    \item (easy to use) This system was easy to use.
\end{enumerate}

\subsection{Final Survey Results}
\label{sec:appendix-finalsurvey}

After experiencing both conditions, participants answered the following preference questions:

Additionally, participants rated the utility of specific Treadstone views (Feed, Branch, Activity, Summary, Data) on a 5-point scale.


The high ratings for the Feed view ($M=4.50, SD=0.78$) and Data view ($M=4.08, SD=1.18$) support $G1$ (Shared Workspace), as participants noted that having all artifacts and dialogues within a single timeline reduced the cognitive cost of context switching (\autoref{sec:qualitative} Finding 4). 
Regarding $G2$ (Transparent and Traceable Contributions), the Summary view ($M=3.96, SD=1.07$) proved effective for synthesizing key findings, while the Branch view ($M=3.67, SD=1.49$) and Activity Tab ($M=3.60, SD=1.35$) provided critical structural grounding to help users reorient within parallel explorations. 
Participant feedback confirmed that \toolname achieved $G3$ (Lightweight Steering) through the banner-style recommendations and the \textbf{Like} button, which were described as convenient tools that suggested analytical directions when they felt stuck. Finally, the system supported $G4$ (Diverse Contributions) by allowing analysts to filter contributions by type within the Activity View, providing what one user described as a \textit{``much wider breadth of perspective''} (P8).

\subsection{Semi-Structured Interview Questions}
\label{sec:appendix-interview}

The final interview explored participants' reasoning behind their survey responses.

\begin{enumerate}
    \item What do you think was the biggest difference between ChatGPT and Treadstone?
    \item Which view or feature of Treadstone was the most useful? Why?
    \item Which view or feature of Treadstone was the least useful or most in need of improvement? Why?
    \item What are the advantages and disadvantages of analyzing data with multiple agents through Treadstone's feed interface, compared to a 1:1 chatbot (e.g., ChatGPT, Claude)?
    \item How intuitive and effective was the approach of using `likes' and suggestion banners to guide further analysis, instead of directly typing prompts?
    \item How helpful were the Activity Tab and Branch View in understanding the overall context of the analysis?
    \item Compared to visualizing and analyzing data with ChatGPT, how did your role change when working with the Treadstone agent team?
    \item Were you able to follow the flow of the analysis without losing track when working with multiple agents?
    \item Do you have any additional comments about Treadstone?
\end{enumerate}

\section{Experimental Datasets}
\label{sec:appendix-datasets}

For the exploratory visual analysis tasks in our user study, we utilized two distinct, real-world tabular datasets to prevent learning effects during the counterbalanced sessions. Both datasets offer rich, multi-dimensional attributes suitable for open-ended data exploration.

\subsection{Global Terrorism Database (GTD)}
The GTD is a comprehensive, open-source database of terrorist incidents worldwide. 
\begin{itemize}
    \item \textbf{Size:} Over 180,000 rows (incidents) and 135 columns.
    \item \textbf{Coverage:} Events spanning from 1970 through 2017.
    \item \textbf{Key Attributes:} Date (\textit{year, month, day}), geographic location (\textit{country, region, city, latitude, longitude}), attack details (\textit{attack type, weapon type, success flag}), target information (\textit{target type, nationality}), perpetrator group, and casualty counts (\textit{number killed, number wounded}).
    \item \textbf{Analytical Potential:} This dataset is highly suitable for spatio-temporal trend analysis, comparative studies between regions or attack methods, and identifying historical shifts in global security threats.
\end{itemize}

\subsection{IMDB Top 1000 Movies}
This dataset is a curated collection of 1,000 highly-rated films listed on the Internet Movie Database (IMDB).
\begin{itemize}
    \item \textbf{Size:} 1,000 rows (movies) and 16 columns.
    \item \textbf{Key Attributes:} Title, release year, age certificate, runtime, genre, IMDB rating, meta score, director, top four cast members, number of votes, and gross revenue.
    \item \textbf{Analytical Potential:} This dataset supports a variety of visual analytics tasks, including exploring the correlation between ratings and box office revenue, identifying popular genres over different decades, and analyzing the track records of specific directors or actors.
\end{itemize}

\begin{figure}[htb]
  \centering
  \includegraphics[width=\columnwidth]{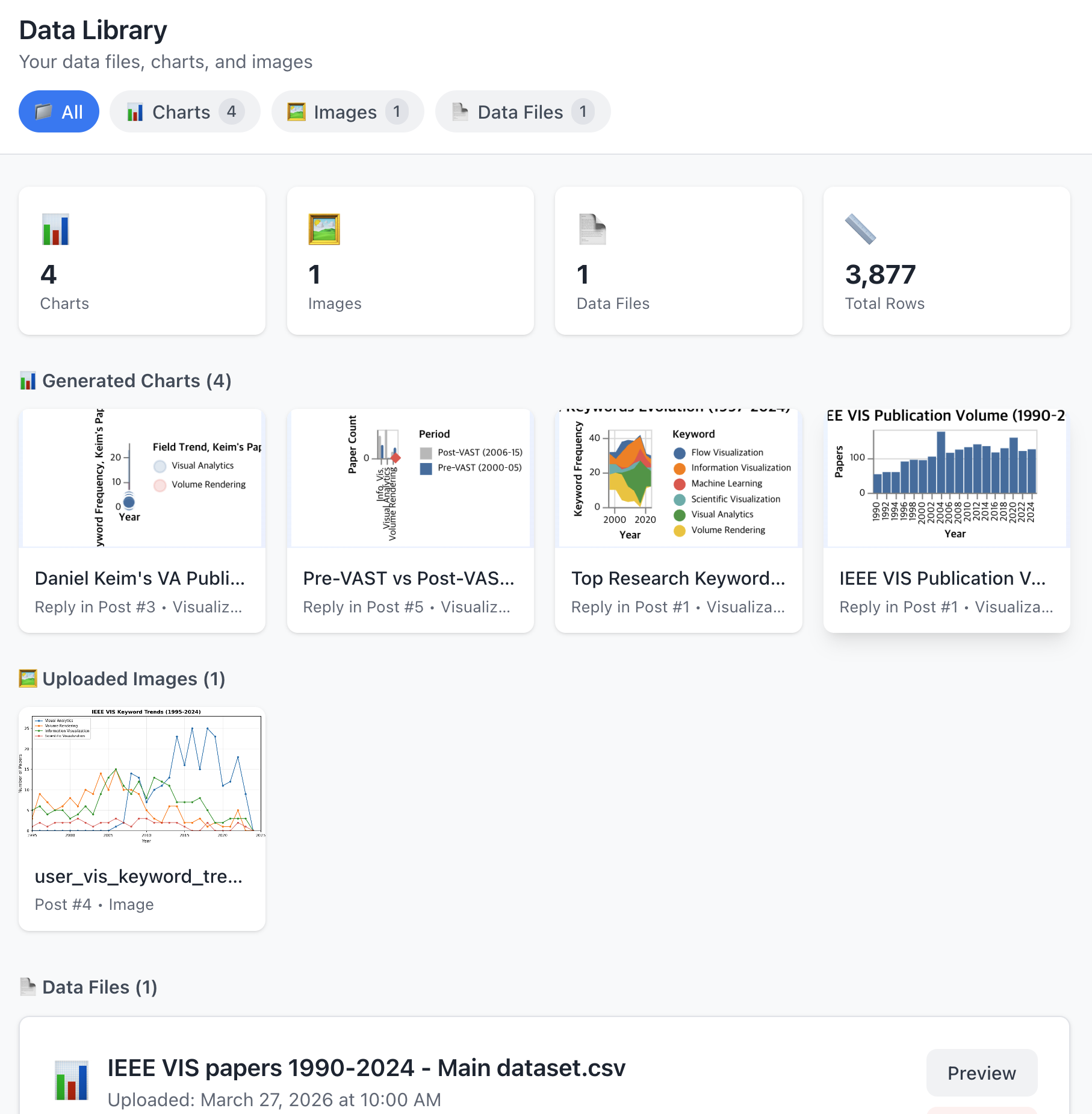}
  \caption{Data View shows analysis artifacts, such as uploaded data and images.  }
  \label{fig:dataview}
\end{figure}

\begin{figure}[htb]
  \centering
  \includegraphics[width=\columnwidth]{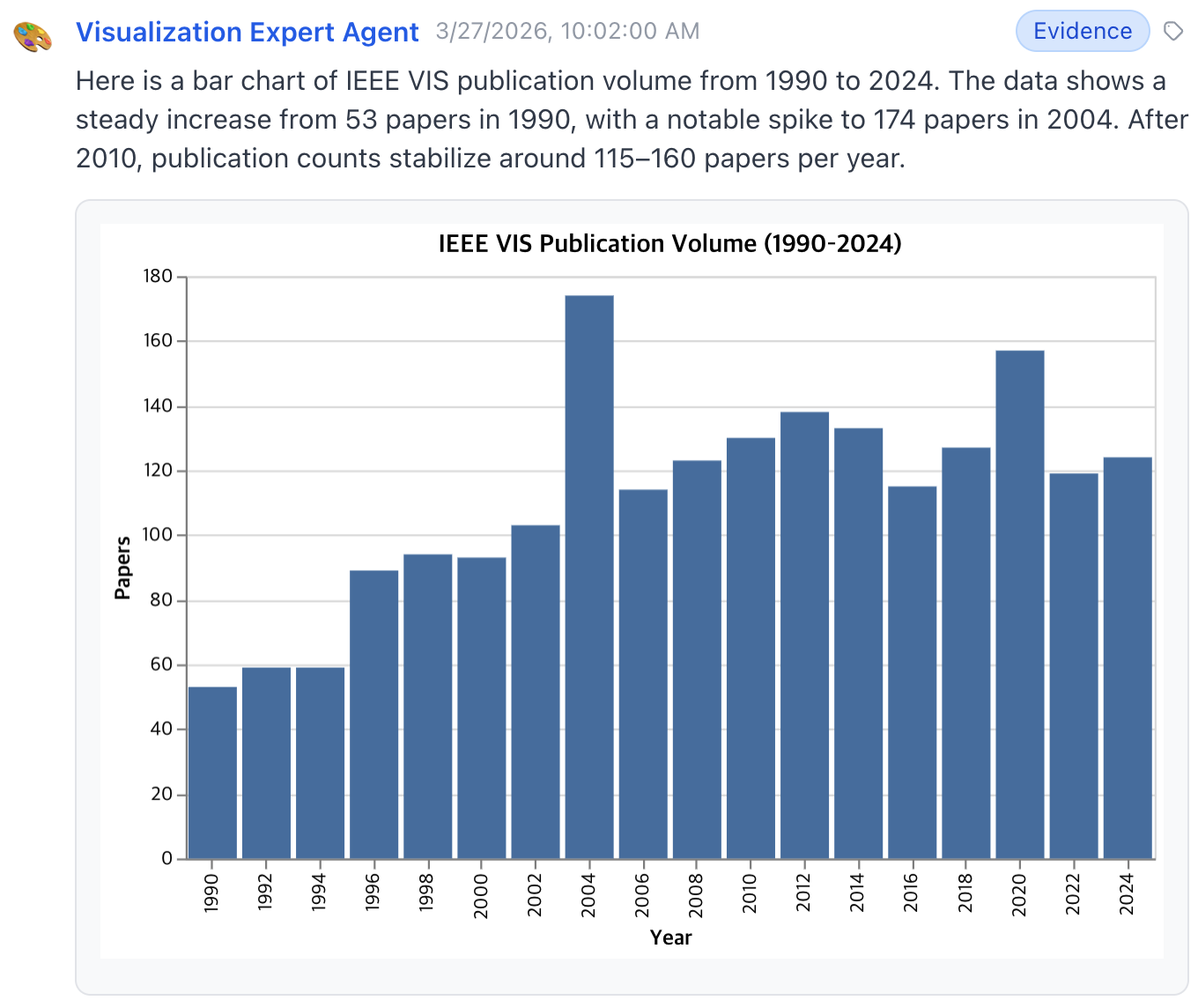}
  \caption{Bar chart generated by the Visualization Expert agent in the usecase scenario. }
  \label{fig:usecasebarchart}
\end{figure}


\begin{figure}[htb]
  \centering
  \includegraphics[width=\columnwidth]{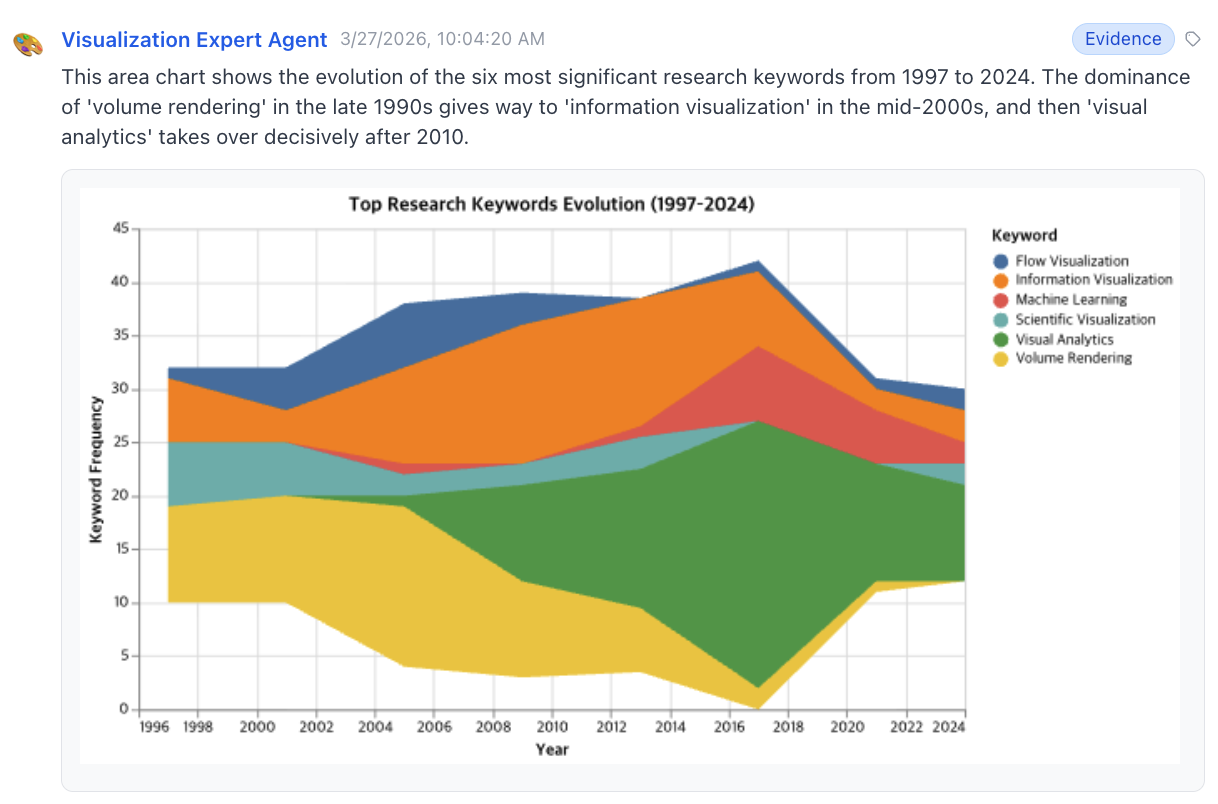}
  \caption{A stacked line chart for keyword trend analysis created by the visualization expert agent. }
  \label{fig:chartfromvizagent}
\end{figure}

\end{document}